\documentclass[a4paper,fleqn,usenatbib]{mnras}
\usepackage{newtxtext,newtxmath}
\usepackage[T1]{fontenc}
\usepackage{ae,aecompl}
\usepackage{graphicx}
\usepackage{amsmath}
\usepackage{bm}
\usepackage{physics}
\usepackage{soul}

\newcommand{\pmbu}{\pmb{u}}

\newcommand\ab{\bm{a}}

\newcommand\rb{\bm{r}}
\newcommand\Db{\bm{D}}
\newcommand\mub{\bm{\mu}}
\newcommand\sigmab{\bm{\sigma}}
\newcommand\thetab{\bm{\theta}}

\title{Mass, Morphing, Metallicities: The Evolution of Infalling High Velocity Clouds}
\author[Heitsch et al.]
{\parbox{\textwidth}{F. Heitsch,$^{1}$\thanks{E-mail: fheitsch@unc.edu} A. Marchal,$^{2}$ M.-A. Miville-Desch\^{e}nes,$^{3}$ J.~M. Shull,$^{4}$ A.~J. Fox,$^{5}$ 
}\vspace{0.4cm}\\
\parbox{\textwidth}{
$^{1}$Department of Physics and Astronomy, University of North Carolina Chapel Hill, Chapel Hill, NC 27599-3255, U.S.A\\
$^{2}$Canadian Institute for Theoretical Astrophysics, University of Toronto, Toronto, ON M5S 3H8, Canada\\
$^{3}$Laboratoire AIM, CEA-Saclay, CNRS, Universit\'{e} Paris-Saclay, Universit\'{e} Paris Diderot, Sorbonne Paris Cit\'{e}, F-91191 Gif-sur-Yvette, France\\
$^{4}$CASA, Department of Astrophysical and Planetary Sciences, University of Colorado Boulder, CO 80309, U.S.A\\
$^{5}$AURA for ESA, Space Telescope Science Institute, 3700 San Martin Drive, Baltimore, MD 21218, U.S.A}}

\date{Accepted XXX. Received YYY; in original form ZZZ}

\pubyear{2018}

\begin{document}
\label{firstpage}
\pagerange{\pageref{firstpage}--\pageref{lastpage}}
\maketitle

\begin{abstract}
We revisit the reliability of metallicity estimates of high velocity clouds with the help of hydrodynamical simulations. We quantify the effect of accretion and viewing angle on metallicity estimates derived from absorption lines. Model parameters are chosen to provide strong lower limits on cloud contamination by ambient gas. Consistent with previous results, a cloud traveling through a stratified halo is contaminated by ambient material to the point that $<10$\% of its mass in neutral hydrogen consists of original cloud material. Contamination progresses nearly linearly with time, and it increases from head to tail. Therefore, metallicity estimates will depend on the evolutionary state of the cloud, and on position. While metallicities change with time by more than a factor of $10$, well beyond observational uncertainties, most lines-of-sight range only within those uncertainties at any given time over all positions. Metallicity estimates vary with the cloud's inclination angle within observational uncertainties. The cloud survives the infall through the halo because ambient gas continuously condenses and cools in the cloud's wake and thus appears in the neutral phase. Therefore, the cloud observed at any {\em fixed} time is not a well-defined structure {\em across} time, since material gets constantly replaced. The thermal phases of the cloud are largely determined by the ambient pressure. Internal cloud dynamics evolve from drag gradients caused by shear instabilities, to complex patterns due to ram-pressure shielding, leading to a peloton effect, in which initially lagging gas can catch up to and even overtake the head of the cloud.
\end{abstract}


\section{Background}
The Galactic halo hosts a population of clouds whose line-of-sight velocities are inconsistent with Galactic rotation \citep{1997ARA&A..35..217W}. Historically identified by their neutral gas component observed via H{\small{I}}~$21$~cm emission, their mass budget is actually dominated by ionized gas \citep{2004ApJ...602..738F,2009ApJ...699..754S,2011Sci...334..955L}. These high velocity clouds (HVCs) range from large complexes of many degrees to structures at the resolution limit. Their diversity suggests different origins \citep{1997ARA&A..35..217W,2012ARA&A..50..491P}.

HVC metallicities may provide key information about cloud origins. Metallicities range between $10-50$\% solar
\citep{1999Natur.402..388W,2010ApJ...718.1046F,2011ApJ...739..105S,2013ApJ...772..111R,2007ApJ...657..271C,2017ASSL..430...15R},
possibly indicating satellite or circum-Galactic material as an origin. Intermediate and low velocity clouds (IVCs, LVCs) have higher metallicity, suggesting a Galactic wind source \citep{2012ARA&A..50..491P}. Yet, mixing with or accretion of ambient material affects the reliability of metallicity constraints as indicators of HVC origin \citep{2014ApJ...795...99G,2015ApJ...812..111K,2016MNRAS.462.4157A,2016ApJ...816L..11F,2017ApJ...837...82H}.

The contamination problem is one of turbulent mixing \citep{2006ApJ...648.1043E,2010ApJ...719..523K}. Yet, turbulent mixing is a challenging problem, especially in inviscid hydrodynamic simulations. Convergence is hard to reach, and mixing can be dominated by the numerical scheme \citep{2008ApJ...680..336S,2017MNRAS.468.3184G} and by the physics included in the modeling \citep{1997ApJ...483..262V,2007A&A...472..141V,2015MNRAS.449....2M,2021MNRAS.501.5330S}. More recently, the focus has shifted to contamination by accretion relying on cooling. A sizeable fraction of the HVC population may be Galactic fountain material \citep[][see also \citealp{2012ARA&A..50..491P}]{2015MNRAS.447L..70F,2017MNRAS.464L.100M}. Gas launched from the disk generates a wake, leading to condensation of halo gas due to radiative losses \citep[][see also \citealp{2009ApJ...698.1485H}]{2020MNRAS.492.1970G}. Up to $100$\% of the original cloud mass may be accreted from the ambient gas \citep{2017ApJ...842..102G}.

Best suited as a laboratory for HVCs are the Smith cloud \citep{1963BAN....17..203S,1998MNRAS.299..611B,2008ApJ...679L..21L}, and Complex C \citep{1966BAN....18..413H}. Distances and orientations are well constrained (\citealp{2003ApJS..146....1W,2003ApJ...597..948P,2008ApJ...679L..21L} for the Smith Cloud; \citealp{2007ApJ...670L.113W,2008ApJ...684..364T} for Complex C), to the point that trajectory reconstructions have been attempted \citep{2014MNRAS.442.2883N,2015MNRAS.447L..70F}. \citet{2017ApJ...837...82H} compare their simulation results with the Smith Cloud, exploring different evolution times and viewing angles. They conclude that the probability to find "original" cloud material decreases with distance from the leading head of the cloud, and that therefore the origin of the Smith Cloud remains uncertain. 

We revisit the contamination scenario, focusing on the probability to identify original cloud material in the neutral gas phase. Our analysis differs in four aspects from \citet{2017ApJ...837...82H}; (1) Instead of a wind-tunnel experiment, the cloud is dropped in a Galactic gravitational potential: (2) The analysis of "original" cloud content is generalized to arbitrary viewing angles; (3) Column densities for metallicities are derived from line ratios, which in turn are determined by a Gaussian decomposition of the emission and absorption line profiles using {\texttt{ROHSA}} \citep{2019A&A...626A.101M}. (4) We allow the gas to cool below $10^4$~K. Radiative losses at $10^5$~K already lead to cooling-induced fragmentation \citep{2020ApJ...903..101S}. Cooling below $10^4$~K can lead to the formation of cold neutral gas via thermal instability \citep{1965ApJ...142..531F,1995ApJ...443..152W,1995ApJ...453..673W}, and thus can enhance overall fragmentation of the cloud \citep[][for this effect on Galactic wind models]{2016ApJ...821....7T}.

We generated a set of hydrodynamic models of HVCs traveling through a diffuse halo medium (Sec.~\ref{s:modeldesc}). We assess the probability to find cloud material at arbitrary positions (Sec.~\ref{ss:probhvchighmass}) and estimate the effects of contamination and viewing angle on abundance estimates (Sec.~\ref{ss:timeangle}). We explore correlations between centroid velocities and metallicities (Sec.~\ref{ss:lagvelo}) and between column densities and metallicities (Sec.~\ref{ss:colmetal}) as indicators for contamination. Caveats and consequences for observational estimates are mentioned in Sec.~\ref{s:discussion}.


\section{Model Description}\label{s:modeldesc}
We use a modified (App.~\ref{s:athenamod}) version of Athena~4.2 \citep{2008ApJS..178..137S}, an Eulerian (fixed) grid code solving the equations of inviscid, compressible hydrodynamics in conservative form,
\begin{eqnarray}
\frac{\partial \rho}{\partial t}+\nabla\cdot(\rho\pmbu)=0\label{eq:mass}\\
\frac{\partial \rho C_c}{\partial t}+\nabla\cdot(\rho C_c \pmbu)=0\label{eq:tracercloud}\\
\frac{\partial \rho C_h}{\partial t}+\nabla\cdot(\rho C_h \pmbu)=0\label{eq:tracerhalo}\\
\frac{\partial \rho\pmbu}{\partial t}+\nabla\cdot(\rho\pmbu\pmbu+P)=-\rho\nabla\Phi\label{eq:momentum}\\
\frac{\partial E}{\partial t}+\nabla\cdot\left((E+P)\pmbu\right)=\rho\mathfrak{L}\label{eq:energy},
\end{eqnarray}
with the total mass density $\rho$, the gas velocity $\pmbu$, the pressure $P$ and the total energy
\begin{equation}
  E\equiv\frac{1}{2}\rho\pmbu^2+\frac{P}{\gamma-1}.\label{eq:totener}
\end{equation}
The "dye" for the cloud ($C_c$) [the halo ($C_h$)] is initialized to $1$ wherever there is cloud [halo] material, and to $0$ elsewhere. The adiabatic exponent is set to $\gamma=5/3$, and the mean molecular weight to $\mu=1$ in units of $m_{\rm H}$. The latter choice affects cloud masses and sound speeds. Therefore, we will quote mass ratios, and any time-scales are approximations. The effective equation of state is determined by a combination of heating and cooling processes $\mathfrak{L}$ (App.~\ref{s:thermalphysics}) applied to the total energy equation (Eq.~\ref{eq:totener}). We use the HLLC solver \citep{1994ShWav...4...25T} in combination with a third-order Runge-Kutta integrator  (App.~\ref{ss:timestepping}).

\subsection{Initial and Boundary Conditions}\label{ss:modeltypes}
Not much is known about the "initial" shape and structure of HVCs, if such a concept is valid at all. A popular choice  is to assume a spherical cloud with some density and temperature profile \citep[e.g.][but see \citealp{2009ApJ...703..330C}, though in a different context]{2001ApJ...555L..95Q,2007A&A...472..141V,2009ApJ...698.1485H,2014ApJ...795...99G,2021MNRAS.501.5330S}, which is then exposed to a wind, or, in our case, is falling through the Galactic halo. Such a setup emphasizes the cloud survival aspect of a HVC simulation. This is not surprising, since many previous studies using such initial conditions were interested in the survival of original cloud material. 

We approach the choice of initial conditions differently, since uniform-density initial conditions can leave a clear imprint on the cloud structure for a substantial time \citep{2017ApJ...837...82H}.  We do not aim at setting up a realistic cloud in the initial conditions, but we develop initial conditions that evolve into a realistic cloud, and take that cloud as our starting point for the analysis. The reader could think of this as a burn-in phase; the first half of the simulation ($\sim 140$~Myr) is used to set up the cloud. During this phase, the cloud mostly preserves its initial structure and consists of original cloud material, with only minor contributions from ambient gas. Since we are considering an idealized numerical experiment investigating the contamination of a model cloud by ambient gas, it does not matter whether the cloud formed via stripping of gas from dwarf satellites \citep{2011ApJ...742..110N}, via Galactic winds/outflows \citep{2015MNRAS.447L..70F}, or via any other scenario.

In this spirit, we imagine that the HVC starts its voyage at an initial distance of $\sim 30$~kpc above the plane, providing the cloud with sufficient room to lose any imprint of the initial conditions. We encase the cloud in a simulation domain such that it rests in the lower eighth of the box. The box measures $1\times 1\times 8$~kpc$^3$ with $256$ cells along the short dimension. The box is three times as wide as the initial cloud diameter, which is resolved by $85$ cells. This should suffice to capture the large-scale dynamics of the cloud in three dimensions \citep{2016MNRAS.457.4470P,2017MNRAS.468.3184G}. We set an initial cloud mass of $M_c= 5\times 10^6$~M$_\odot$, comparable to that of Complex C \citep[][see also \citealp{2007ApJ...670L.113W}]{2008ApJ...684..364T} or -- depending on the distance estimate -- the Smith Cloud \citep{1998MNRAS.299..611B,2008ApJ...679L..21L}. The latter authors arrive at a lower mass limit for H{\small{I}} of $10^6$~M$_\odot$ based on a comparison of several distance estimates placing the cloud at $12.4\pm 1.3$~kpc. The cloud is chosen to be sufficiently massive so that some of its neutral gas survives the passage through the halo \citep{2009ApJ...698.1485H}. The {\em initial}, average cloud density is $n_H\approx 10$~cm$^{-3}$, or a factor of $10^2\cdots 10^3$ higher than the average HVC density for Complex C \citep{2008ApJ...684..364T,2011ApJ...739..105S}. Once we start the analysis, peak cloud densities have dropped to $\sim1$~cm$^{-3}$, with average densities of $\sim0.1$~cm$^{-3}$.

The cloud itself is realized via a double top-hat density profile, with temperatures corresponding to the two-phase equilibrium temperatures (see Appendix~\ref{s:thermalphysics} for details on the thermal physics). We identify neutral gas (H{\small{I}}) via the ionization fraction $x_i$ that is implicitly calculated as a function of total density $n_{\rm tot}$ in our thermal physics description. Thus, the neutral column density is $N$(H{\small{I}})$=\int n_{\rm tot}(s)(1-x_i(s))ds$ along a line of sight $s$. Initially at rest, the cloud does not have an ionized gas component, but this component will develop during the burn-in phase, reaching $50$\% of the total cloud mass. 

A fixed gravitational potential accelerates the cloud toward the disk. The potential consists of a combination of dark matter halo, spheroidal bulge and disk potentials  \citep{2000MNRAS.314..511S,2008ApJ...674..157C,2016ApJ...821....7T}. We assume a Galacto-centric radius of $8.5$~kpc, compared to $7.6$~kpc for the Smith Cloud \citep{2008ApJ...679L..21L}. For simplicity regarding boundary conditions, we let the cloud drop perpendicularly to the Galactic plane. We do not assume a dark matter halo confining the cloud \citep{2001ApJ...555L..95Q,2009ApJ...707.1642N,2016ApJ...816L..18G}, nor do we include self-gravity \citep{2021MNRAS.501.5330S}.

We place the cloud in an initially isothermal, hydrostatically stratified halo determined by the gravitational potential, with $T=2.2\times 10^6$~K \citep{2012ApJ...756L...8G,2013ApJ...770..118M,2015ApJ...800...14M}. \citet{2014ApJ...784...54H} discuss the justification for {\em non}-isothermal models, at lower temperatures. Our choice of an isothermal model is motivated by numerical considerations of stability. Non-isothermal models are prone to convection \citep{2014ApJ...784...54H}, and in combination with lower assumed temperatures \citep{2012arXiv1211.4834W} would require additional mechanical heating mechanisms such as galactic winds or cosmic rays \citep{2016ApJ...816L..19G} to achieve the appropriate scale heights. The required densities ($n\approx  10^{-4}$~cm$^{-3}$ at $z=30$~kpc) in our model halo are low enough that the cooling time is longer than the typical crossing time through the simulation box, and therefore, the ambient gas temperature does not change drastically as long as the gas is not perturbed by the cloud's passage. Lower halo temperatures would result in stronger cooling and thus in faster contamination of the cloud by ambient halo gas. We neglect any possible co-rotation of the warm halo gas with the Galactic disk \citep{2020ApJ...894..142Q}. The cloud is initially in thermal pressure balance with the halo, at a pressure of $P/k_B=660$~K~cm$^{-3}$. We apply a grid of velocity perturbations drawn from a turbulent spectrum with index $|v|^2\propto k^{-4}$ on a wave number range $1/r_\mathrm{c}\leq k \leq (2\Delta x)^{-1}$, where $r_\mathrm{c}$ is the cloud radius. The specific choice of the turbulent spectrum does not affect the results. The velocity perturbations are normalized such that they generate a turbulent pressure $\rho \sigma^2/k_B=330$~K~cm$^{-3}$, or $50$\% of the thermal pressure. The velocity perturbations break the initial symmetry of the setup, suppressing numerical artifacts typical for inviscid solvers in the presence of highly symmetric initial conditions. The kinetic energy from the turbulent motions make the cloud expand during the burn-in phase, pushing some cloud gas from the neutral into the ionized component (Sec.~\ref{ss:probhvchighmass}).

We assume solar metallicity ($Z/Z_\odot=1$) for the cloud, and $Z/Z_\odot=10^{-3}$ for the ambient gas. \citet{2015ApJ...800...14M} infer $Z/Z_\odot=0.3$ for the halo, and metallicities for the Smith Cloud and Complex C are estimated at $Z/Z_\odot=0.53^{+0.21}_{-0.15}$ \citep{2016ApJ...816L..11F} and $Z/Z_\odot=0.1-0.3$ \citep{2007ApJ...657..271C}, respectively. Therefore, our choices are not motivated by observations, but rather by an attempt (a) to clearly distinguish between cloud and halo material, and (b) to provide strict lower limits for cloud contamination estimates. Choosing metallicities closer to observed values would make it harder to distinguish between cloud and halo material, and higher halo metallicities would lead to faster cooling and thus more accretion of halo gas entrained in the cloud's wake. Taken together with the assumed halo temperature, our model cloud is therefore {\em less likely} to be contaminated than observed clouds. Even with these extreme choices, distinguishing between cloud and ambient material proves to be a challenge.

Cloud and halo gas are identified by passively advected scalar fields that can be imagined as a kind of dye being evolved together with the hydrodynamical equations (eqs.~\ref{eq:tracercloud}, \ref{eq:tracerhalo}). The cloud mass at any time is given by 
\begin{equation}
  M_\mathrm{c}=\int \rho(\mathbf{x})C_\mathrm{c}(\mathbf{x})\,d^3x,\label{e:defcloud}
\end{equation}
with a color field $C_\mathrm{c}$ that is initialized to $1$ within the cloud and $0$ otherwise. The gas density is given by $\rho=\mu m_\mathrm{H} n$, where we set $\mu=1$. Similarly, we identify original halo material, using a tracer field $C_\mathrm{h}$. 

Following \citet{2008ApJ...680..336S} and \citet{2017MNRAS.468.3184G}, we integrate the hydrodynamical equations within the rest frame of the cloud (App.~\ref{s:comovinggrid}), to improve the numerical accuracy. Since the cloud moves downward within the simulation box, the upper and lower boundary conditions in the vertical direction need to be modified. For the lower boundary condition, we calculate the hydrostatic density at the current vertical position of the boundary, and feed the corresponding hydrodynamic quantities to the boundary cells. The upper boundaries are set to 'open if leaving'.

\subsection{Data Analysis}
Instead of post-processing, we have Athena generate position-position-velocity (PPV) cubes at $9$ inclination angles from $10\cdots 90^\circ$ on the fly (App.~\ref{ss:ppvcubes}). The cubes contain H{\small{I}}-21~cm and optically thin generic metal absorption line spectra at each position. The metallicities can change drastically within the cloud. An average metallicity can be derived by adding up column densities along the line of sight. To get a more accurate measure of the metallicity at each position, we mimic the observational procedure to estimate column densities for individual velocity components for both H{\small{I}}-21~cm and the generic metal tracer. We decompose all spectra ($\sim 6\times 10^4$ spectra per time step, resulting in a total of $\sim 4.5\times10^5$ spectra) into Gaussian components using {\texttt{ROHSA}} \citep[][see App.~\ref{s:ROHSA} for details]{2019A&A...626A.101M}.


\section{Results}\label{s:results}


\subsection{Raw Data and Overall Evolution}

\begin{figure*}
  \centering
  \includegraphics[width=\linewidth]{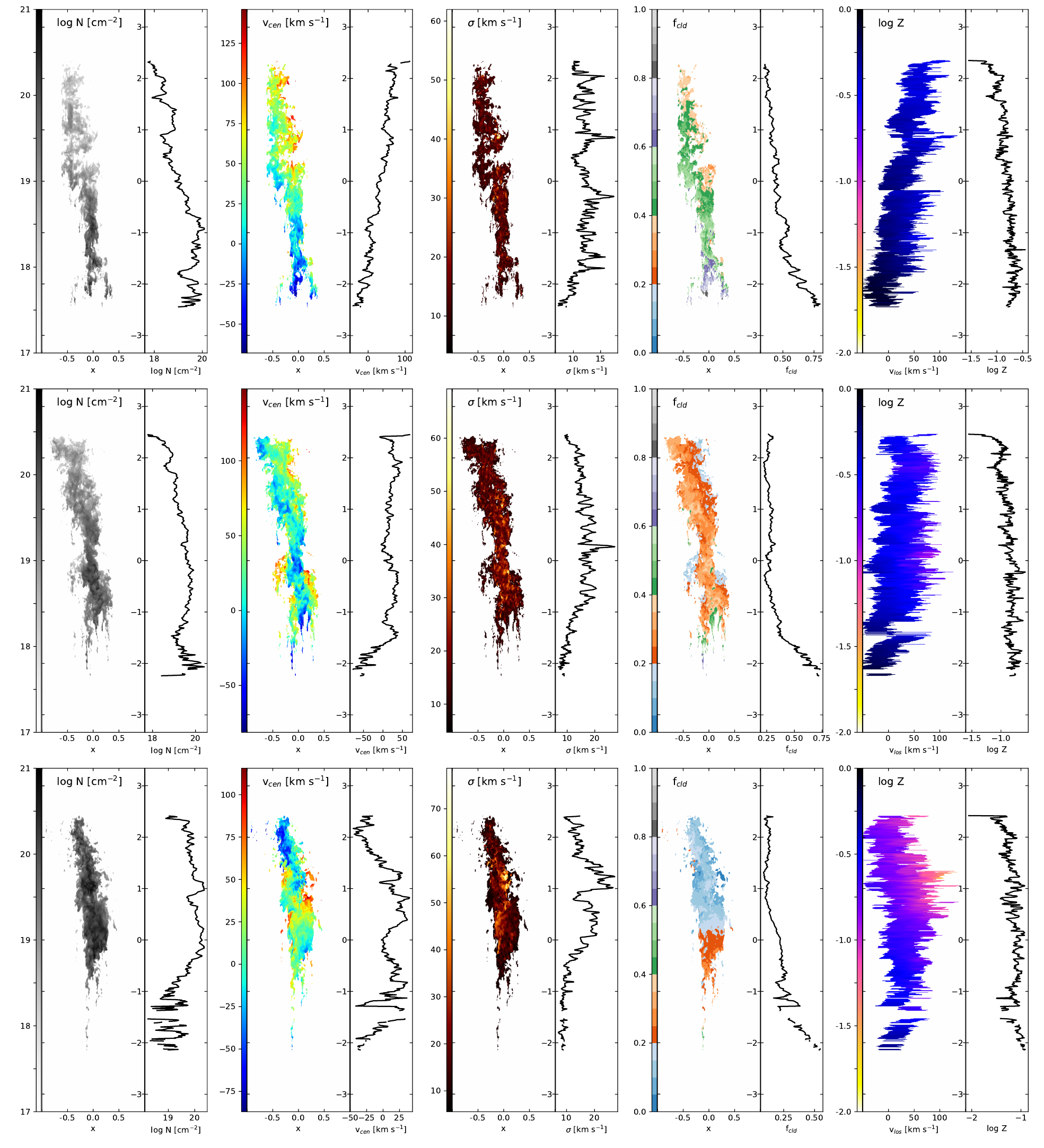}
  \caption{\label{f:allquant}Integrated maps of the model cloud traveling through the halo toward the Galactic plane, at $200$, $250$, and $300$~Myr (compare to Fig.~\ref{f:probtime}). The cloud is viewed under an inclination angle of $\alpha_i=50^\circ$. At $\alpha_i=90^\circ$  the cloud moves toward the observer. Each map is accompanied by its profile along the vertical axis. The spatial extent is given in kpc on the map's abscissa and the profile plot's ordinate. Color bars show quantities indicated on the profile plot's abscissa. From left to right: Neutral (H{\small{I}}) column density, centroid velocity (eq.~[\ref{e:vcen}]), total line width (eq.~[\ref{e:vsig}]), cloud mass fraction, and metallicity in terms of centroid velocity and position along the cloud's long axis (see text). The first three quantities are calculated from the {\tt ROHSA} component fitting (Sec. \ref{subsec:estimating-parameters}) of H{\small{I}}-$21$~cm emission, and the last two quantities are based on the simulation raw data. At early times, column densities, centroid velocities, and metallicities show gradients, while at late times, the dynamics and hence the maps get more complex.}
\end{figure*}

The cloud passes through the stages typical for HVC simulations, from a flattened appearance   \citep[see][for a demonstration]{2017ApJ...837...82H,2021MNRAS.501.5330S} to an elongated head-tail structure. The cloud develops a diffuse, shock-compressed halo and a dense, small core. These are transients arising from the initial conditions, and they decay quickly to values consistent with the ambient pressure, leading to a more extended cloud that develops a tail (Fig.~\ref{f:allquant}, top row). We skip those early stages in our analysis, since they are nearly completely determined by the choice of initial conditions \citep{2009ApJ...703..330C}. Ram pressure decelerates the outer, more diffuse material, resulting in differential drag \citep{2007ApJ...656..907P} and therefore in a centroid velocity gradient. The cloud begins to stretch out. Buoyancy effects in the stratified halo may start to play a role. At this point, a combination of radiative cooling and hydrodynamic instabilities fragments the cloud (Fig.~\ref{f:allquant}, center row). The cloud hits the Galactic plane at ${\rm v}_0\sim 300$~km~s$^{-1}$, close to the ballistic velocity, and in approximate agreement with the analytic estimates of \citet{2020ApJ...903..101S}.
We discuss the amplitude of the impact velocity further in Sec.~\ref{sss:impactvel}.

\subsubsection{Trajectories}
\begin{figure}
    \includegraphics[width=\columnwidth]{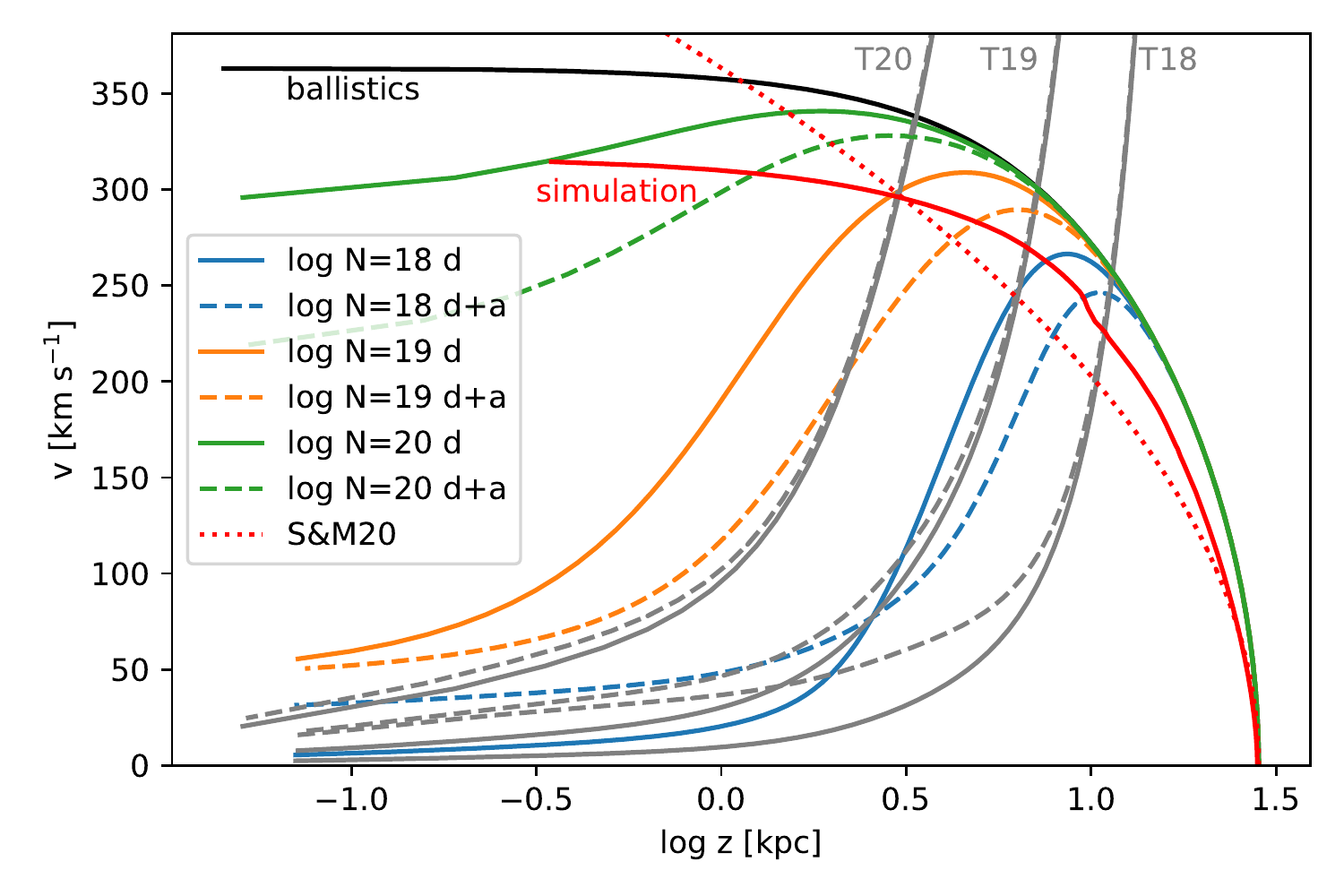}
    \caption{Phase space trajectories following \citet{1997ApJ...481..764B} for ballistic clouds, compared to the hydrodynamical models (red curve). Blue, orange, and green trajectories stand for clouds at the indicated column densities. We compare models with drag only (solid lines), and with drag and accretion (dashed lines). Local terminal velocities \citep[][their eq.~2]{1997ApJ...481..764B} are shown in grey for three column densities (labels T18, T19, T20). With increasing cloud column, drag is less efficient in decelerating the cloud. The red dotted line indicates the analytic estimate by \citet{2020ApJ...903..101S} using a logarithmic potential.}
    \label{f:trajectories}
\end{figure}

Figure~\ref{f:trajectories} compares the model cloud trajectory with that of test particle clouds, following the approach of \citet[][see their Fig. 3]{1997ApJ...481..764B}. We extend their discussion by considering mass accretion in addition to drag forces. The trajectories are calculated for three column densities ($10^{18},10^{19},10^{20}$~cm$^{-2}$), including drag (label 'd') and both drag and mass accretion (label 'd+a'). The underlying equations are 
\begin{eqnarray}
  \dot{{\rm v}}&=&-g(z) + \left(\frac{1}{2}C_{\rm d}+1\right){\rm v}\,\dot{m_l}\\
  \dot{m_l}&=&C_{\rm a}\frac{n_{\rm h}(z)}{N_{\rm c}}{\rm v},
\end{eqnarray}
with $\dot{z}\equiv {\rm v}$. For all models except the red dotted line of \citet{2020ApJ...903..101S}, the gravitational acceleration $g(z)$ is derived from the model halo  (Sec.~\ref{ss:modeltypes}). The ambient halo density is given by $n_{\rm h}(z)$, $C_{\rm d}=1$ is the drag coefficient, $m_l\equiv \ln m$ the logarithmic mass, and $N_{\rm c}$ is the cloud's column density. We quantify the accretion efficiency by an accretion coefficient, $C_{\rm a}=1$.

Drag and accretion result in the same expression, yet headed by different factors. Obviously, only accretion changes the mass. Low-column density clouds approach the terminal velocities \citep[][shown in grey]{1997ApJ...481..764B}, and this is still valid for accreting clouds. The hydrodynamical cloud (red line) does not reach a local terminal velocity, but it approaches a ballistic trajectory. For the first $100$~Myr, the cloud stays fairly coherent (Fig.~\ref{f:probtime}b), dropping nearly ballistically through the halo. The cloud starts to fragment and to accrete material, breaking up the coherent structure. Meanwhile (Fig.~\ref{f:allquant}, Fig.~\ref{f:probtime}[d]), the column density is increasing, due to accretion of ambient material \citep{2010MNRAS.404.1464M,2014ApJ...795...99G,2015MNRAS.447L..70F,2020MNRAS.492.1970G} and because of the increasing ambient pressure \citep{2009ApJ...698.1485H}. For comparison, we also show the cloud velocity estimated by \citet{2020ApJ...903..101S} (red dotted line). This analytic estimate does not include drag, and therefore is expected to over-predict the velocity. In addition, \citet{2020ApJ...903..101S} use a logarithmic potential, which is steeper closer to the disk than our composite potential (Sec.~\ref{s:modeldesc}), but flatter at larger distances.

The cloud is sufficiently massive to punch through the disk, in contrast to the results from models comparing HVCs with and without dark matter halo confinement \citep{2016ApJ...816L..18G}. 
The authors used a similar model setup and parameters as in \citet{2009ApJ...698.1485H} and this paper. The cloud is dropping through a stratified halo, eventually hitting the disk. The initial cloud mass is $5\times 10^6$~M$_\odot$, and the cloud hits the disk at $\sim 300$~km~s$^{-1}$. Yet, \citet{2016ApJ...816L..18G} find that their model cloud  without dark matter confinement cannot punch through the disk. The difference seems to arise from the assumed Galactic disk densities. With $n=0.1$~cm$^{-3}$, our mid-plane density is on the order of the average cloud density. Yet, this mid-plane-density is somewhat low \citep{2001RvMP...73.1031F,2005ARA&A..43..337C}. \cite{2021ApJ...911...55S} find $n=n_{\rm HI}+2n_{H_2}=0.5\cdots 1.5$~cm$^{-3}$ at the solar circle. For their models, \citet{2016ApJ...816L..18G} quote a mid-plane density of $>30$~cm$^{-3}$ (their Fig.~2). 

The simulation stops when cloud material reverses direction, since the role of the boundary conditions would have to be switched. The motions are trans-sonic at this point for a considerable time. Transients interacting with the boundaries start to affect the cloud evolution and thus to invalidate the results.

\subsubsection{Internal Dynamics and Peloton Effect}\label{sss:peloton}
Figure~\ref{f:allquant} shows all quantities reconstructed from the {\tt ROHSA} component fitting (Sec. \ref{subsec:estimating-parameters}) of H{\small{I}}-$21$~cm emission. The cloud is viewed under an inclination angle of $\alpha_i=60^\circ$. At earlier times (top row), column densities, centroid velocities, metallicities, and cloud mass fractions exhibit gradients. After $50$~Myr (center row), the dynamics have turned more complicated and the distinct gradients (top row) have vanished. Of special interest is the structure of the centroid velocity, which starts developing inversions. Clumps of blue-shifted (faster) material appear over the whole length of the cloud. These inversions can also be seen in the associated vertical profiles. At early times, the fastest material is leading. Gas directly behind the cloud head is shielded from the ram pressure of the ambient medium \citep[see also][]{2019AJ....158..124F}. Radiative losses lead to condensation, so that the trailing gas contracts and -- still shielded from ram pressure -- can catch up to the leading head, eventually overtaking it. In other words, gas parcels within the cloud start to switch positions, not unlike in a peloton during bicycle races. 

The line-of-sight velocity dispersion $\sigma_{\rm los}$ (3rd column of Fig.~\ref{f:allquant}) is generally smaller in the leading part of the cloud, indicating that these regions are less turbulent. This is consistent with some observations of HVCs \citep{2001A&A...370L..26B,2007ApJ...656..907P}, as are the gradients in the centroid velocity.  At later times, the cloud starts to heat up dynamically and gets more turbulent. We calculate $\sigma_{\rm los}$ via the second moment of the position-position velocity cube,
\begin{equation}
  \sigma_{\rm los}^2=\frac{\sum N(\rm{v}_{\rm los})(\rm{v}_{\rm los}-\rm{v}_{\rm cen})^2}{\sum N(\rm{v}_{\rm los})},\label{e:vsig}
\end{equation}
where the sum extends over all velocity channels $\rm{v}_{\rm los}$, $N(\rm{v}_{\rm los})$ is the column density for a given velocity channel, and $\rm{v}_{\rm cen}$ is the column-density-weighted centroid velocity 
\begin{equation}
    \rm{v}_{\rm cen}=\frac{\sum N(\rm{v}_{\rm los})\rm{v}_{\rm los}}{\sum N(\rm{v}_{\rm los})}.\label{e:vcen}
\end{equation}
Our calculation of $\sigma_{\rm los}$ contains both thermal and non-thermal contributions to the dispersion. However, at the evolutionary stage of the cloud, the bulk of the gas resides more or less uniformly in the warm regime, around $T=7\times10^3$~K (Fig.~\ref{f:coolingfunc}), corresponding to an isothermal sound speed of $\approx 8$~km~s$^{-1}$. This is smaller than most dispersions shown in the third column of Figure~\ref{f:allquant}, suggesting that gas dynamics, not gas temperatures, drive the variations in $\sigma_{\rm los}$.

\subsubsection{Metallicities and Contamination}

The fourth column of Figure~\ref{f:allquant} shows the cloud mass fraction derived from the simulation data. The cloud mass fraction measures the amount of original cloud material in the observed (neutral) cloud. At early times, the cloud mass fraction indicates original cloud material contaminated by entrained gas along the cloud (grey to orange transition in the cloud mass fraction). With increasing time (center and bottom row), the cloud mass fraction drops, and the cloud consists largely of accreted material. The cloud mass fraction in the center row shows higher (green vs. orange) values at the outskirts close to the leading head, corresponding to the red-shifted centroid velocities. This is again the peloton effect at work, with originally leading material (higher cloud mass fraction, green) being overtaken by newly accreted material (lower cloud mass fraction, orange). 

The fifth column relates the channel velocity ${\rm v}_{\rm los}$ at each position along the cloud's long axis to the metallicity for each channel and position. We generate a position-velocity plot showing the metallicity of each position-velocity pair instead of the intensity. In  other words, each position along the long axis of the cloud is represented by a spectrum where the brightness temperature has been replaced by the channel's metallicity. This information is usually not accessible via observations unless high-resolution spectra are available for both tracers, but it provides us with a detailed view of the correlation between line-of-sight velocity and metallicity. The abscissa shows the channel velocity over the same range as indicated by the colorbar in the centroid velocity maps (2nd column). The ordinate shows, as in all other panels, the position along the cloud's long axis. Color indicates the metallicity for each position and velocity channel. The position-velocity map is derived from the full position-position-velocity cube by integrating over the $x$ (horizontal) axis. At early times (top row), the position-velocity distribution shows a gradient from blue-shifted, leading velocities, to red-shifted, trailing ones. The gradient indicates that the "tail" of the cloud is red-shifted with respect to the head, and therefore lagging behind the head. In the bottom panel, this gradient has vanished, but {\em metallicities} are generally lower at higher velocity for all positions, suggesting contamination. Ambient material at low metallicity is entrained all along the cloud, but it lags behind the main body of the cloud, and therefore travels at more positive (red-shifted) line-of-sight velocities. 

Since most of the cloud material gets replaced by ambient gas, the earlier statement about heating of cloud material should not be taken literally. The velocity dispersion increases mostly because ambient material is entrained in the cloud's wake, then cools and later appears as neutral gas \citep{2020MNRAS.492.1970G}. Actual heating of original cloud material plays only a minor role. While the cloud appears to be heated dynamically, this is just an effect of coherently moving cloud material being replaced by more turbulent ambient material.

 
 \subsection{Probability for Cloud Material}\label{ss:probhvchighmass}
The probability for neutral gas to be cloud material $P(\mathrm{cloud}|\mathrm{HI})$ allows us to assess whether metallicities can constrain cloud origin scenarios. If $z$ is the coordinate along the cloud's long axis, also taken as the local tangent to the trajectory, the probabilities are calculated by determining the ratio of the tracer fields in terms of a mixing parameter averaged laterally over all positions at a fixed $z$,
\begin{equation}
  P(\mathrm{cloud}|\mathrm{HI},z) = \frac{1}{2}\left(1+\frac{\bar{C}_\mathrm{c}(z)-\bar{C}_\mathrm{h}(z)}{\bar{C}_\mathrm{c}(z)+\bar{C}_\mathrm{h}(z)}\right)\label{e:probcloud}
\end{equation} 
where the subscript $c,h$ denote cloud and halo material, respectively. The average color field $\bar{C}_\mathrm{c}(z)$ is defined as
\begin{equation}
  \bar{C}_\mathrm{c}(z) = \frac{\int \rho(\mathbf{x}) C_\mathrm{c}(\mathbf{x}) (1-x_i(\mathbf{x}))\,dx\,dy}{\int \rho(\mathbf{x}) (1-x_i(\mathbf{x}))\,dx\,dy},
\end{equation}
where $x$ and $y$ are the lateral axes, i.e. perpendicular to the trajectory. The local gas density is given by $\rho$, and the ionization fraction is $x_i(\mathbf{x})$. The probability $P(\mathrm{cloud}|\mathrm{HI},z)$ is normalized to $1$ for each $z$.

\begin{figure*}
  \includegraphics[width=\textwidth]{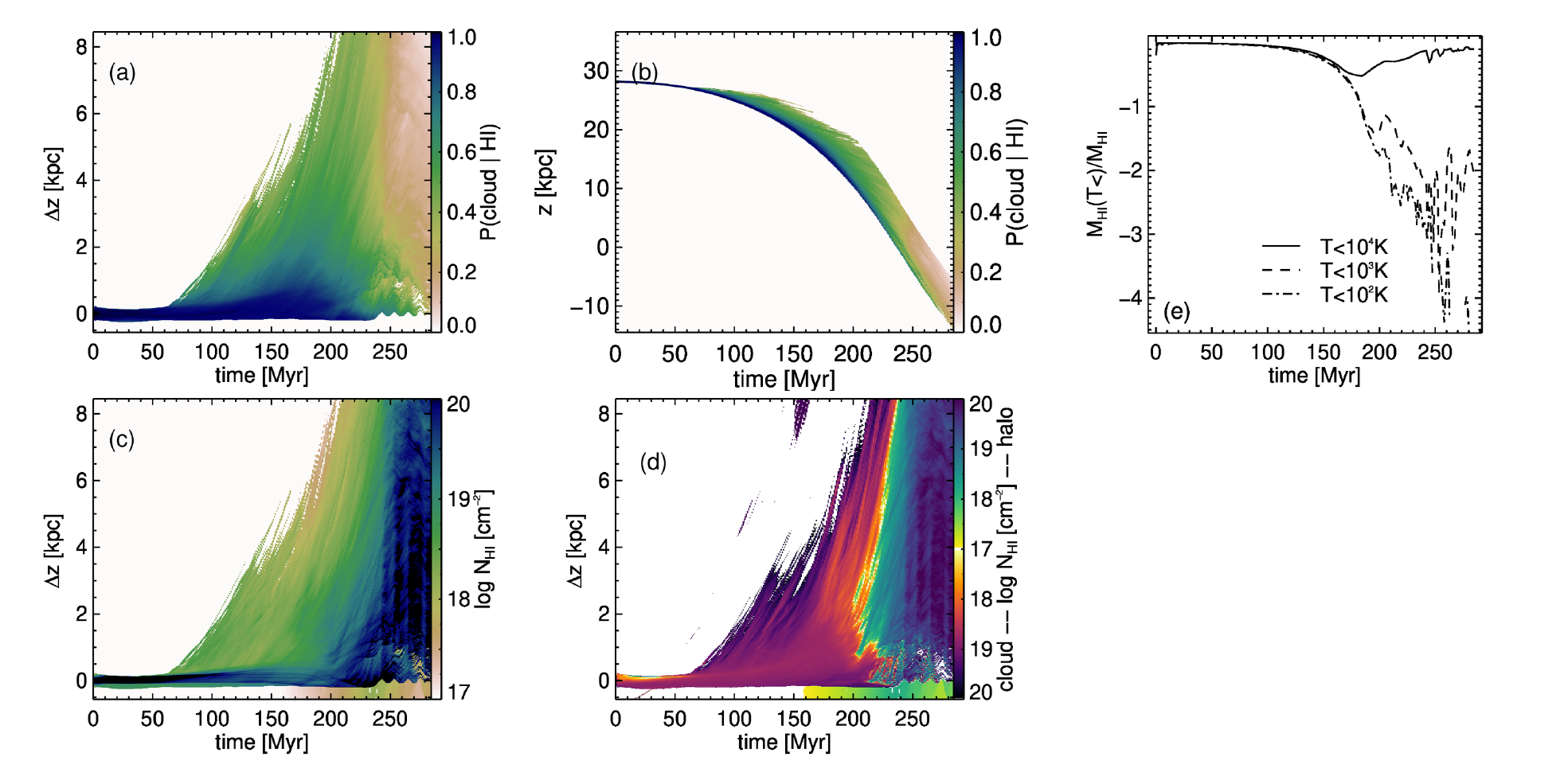}
  \caption{\label{f:probtime}(a) Probability to detect original, neutral cloud gas (eq.~\ref{e:probcloud}) as a function of position along the cloud's major axis and model evolution time. The probability drops to less than 10\% at times $>250$~Myr. (b) Same as (a), but along the cloud's trajectory. (c) Average H{\small{I}} column density along the cloud's long axis, similar to (a). The column density increases toward late times, indicating accretion. (d) H{\small{I}} column density split into contribution from original cloud (purple) and accreted halo (green/blue) material. (e) Mass history for H{\small{I}} at temperatures $T<10^4$~K, $T<10^3$~K and $T<10^2$~K, measuring the gas mass in the warm and cold neutral component. Subsequent analysis is limited to $140<t<240$~Myr, after the cloud has lost the imprint of the initial conditions, and before it hits the Galactic plane.}
\end{figure*}
 
Figure~\ref{f:probtime} summarizes the time evolution of the probability to find cloud material along the long axis of the cloud (panels [a,b]). Initially, the neutral gas component (panel [c,d]) consists of $100$\% cloud material, but the probabilities decrease once a tail forms around $100$~Myr. The cloud ends up mostly consisting of accreted material.  Probabilities $<1$ indicate that ambient material transitioned to the neutral phase via cooling (compare panels [a,c,d]). Its spread along the trajectory  (panel [b]) suggests that this material is entrained in the wake of the cloud, consistent with the metallicity and cloud mass fraction maps of Figure~\ref{f:allquant}. Figure~\ref{f:probtime}e shows the mass of gas at temperatures $\leq 10^4,10^3,10^2$~K, as a measure of the cold gas component. At the end of the burn-in phase, the cold neutral component drops to less than a percent of the neutral cloud mass because of the initial over-pressure within the cloud (\S~\ref{ss:modeltypes}). Once the cloud hits the plane at $t\sim 240$~Myr, the neutral component grows again, via accretion. 

\begin{figure}
    \centering
    \includegraphics[width=\columnwidth]{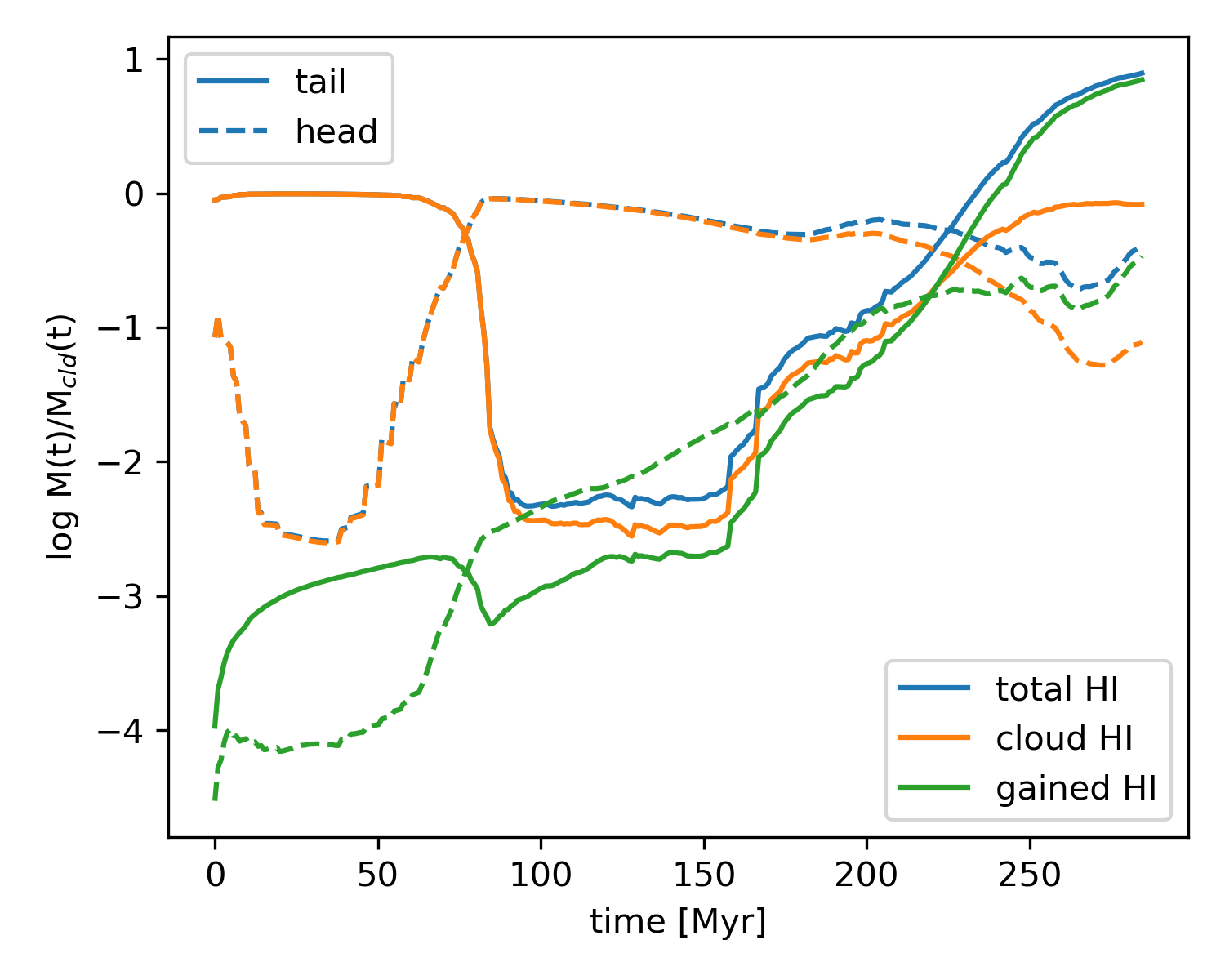}
    \caption{Time evolution of mass fractions over the whole cloud extent (solid) and the leading third (head) of the cloud. Mass fractions refer to the current cloud mass. The cloud tail accretes a factor $\sim 30$ more material than the head. The cloud ends up with $\sim10$\% of original cloud material, both in head and tail.}
    \label{f:masshistory}
\end{figure}

Figure~\ref{f:masshistory} provides a simplified view of the distribution of accreted mass. We split the cloud into a lower third (the 'head' -- dashed lines) and the remaining two thirds (the 'tail' - solid lines). The inversion between head and tail in the total HI gas indicates that initially, the head is heated by compression (dip in mass fraction around $20$~Myr) and then starts to cool again. Meanwhile, the tail loses most of its neutral gas. At around $100$~Myr, the total H{\small{I}}-mass (blue) rises above the cloud H{\small{I}}-mass (orange), a trend that is amplified until the end, when $\sim90$\% of the tail's neutral gas consist of accreted gas. Gained H{\small{I}} (green) supports this interpretation, since it converges on the total H{\small{I}} mass. The mass fraction for the head show a similar evolution, except that they stay a factor of $\sim 30$ below the tail mass fractions at late times. Therefore, most of the gas is accreted in the tail, consistent with \citet{2017ApJ...837...82H}. The mass ratios demonstrate that for our evolved cloud, only $10$\% of the neutral gas is original cloud material. See also Secs.~\ref{ss:timeangle} and \ref{ss:consequences}.

We restrict the subsequent analysis to the time range between $140$ and $240$~Myr, when the cloud has evolved away from its initial conditions, but has not yet crossed the disk's mid-plane.

 \subsection{Thermal Evolution}\label{ss:thermalevol} 
 
 \begin{figure*}
     \centering
     \includegraphics[width=\textwidth]{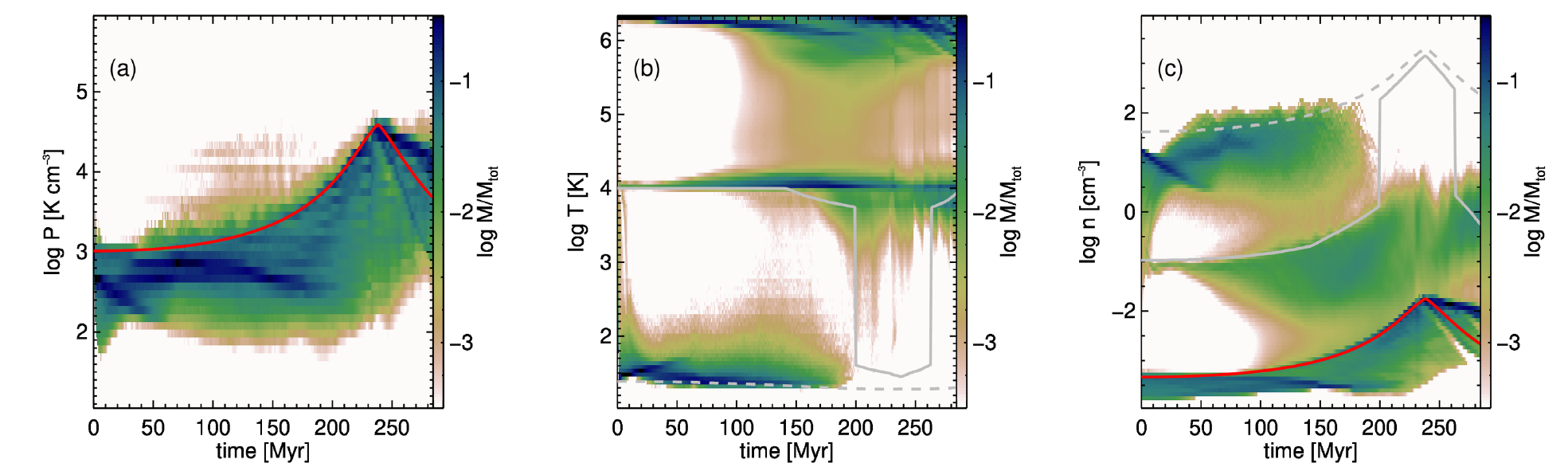}
     \caption{Pressure (a), temperature (b), and density (c) mass fraction distribution against time in the simulation domain. All gas is shown. The ambient halo pressure at the cloud's center of mass (reference position) is shown as a red line in (a), and the corresponding halo density as a red line in (c). The halo temperature is constant at $T=2.2\times 10^6$~K. The grey curves show the thermal equilibrium temperature (b) and density (c) for the dense cloud gas (dashed), and for the cloud enveleope (solid). At higher pressures, there are two solutions, indicating a thermally bi-stable gas (Fig.~\ref{f:coolingfunc}). The ambient pressure drop co-incides with the drop in neutral cloud mass (Fig.~\ref{f:probtime}d).}
     \label{f:thermalevol}
 \end{figure*} 
 
 The neutral gas mass drops again after the cloud punches through the disk (Fig.~\ref{f:probtime}e), suggesting that the ambient thermal pressure may be the driver for the overall mass evolution. Figure~\ref{f:thermalevol} takes a closer look at the thermal state of all gas in the simulation domain. Pressure (Fig.~\ref{f:thermalevol}a), temperature (\ref{f:thermalevol}b) and density (\ref{f:thermalevol}c) distributions in terms of mass fractions are shown against simulation time. Red lines indicate ambient halo values for pressure and density. The bulk of the ambient gas stays at the isothermal halo temperature of $2.2\times10^6$~K, because of the long cooling time. Grey lines in the temperature (Fig.~\ref{f:thermalevol}b) and density (Fig.~\ref{f:thermalevol}c) history show the thermal equilibrium value derived from the cooling curve (Fig.~\ref{f:coolingfunc}) at the current ambient pressure. At the higher pressures around vertical distances of $z=0$ (Galactic plane), two equilibrium solutions are possible , one for warm (solid) and one for cold gas (dashed). The red and grey lines stand for reference values taken at the cloud's center of mass. 
 
 The pressure distribution (Fig.~\ref{f:thermalevol}a) is constrained by the ambient gas pressure for the most part. Ram pressure drives the cloud gas above the red line, and cloud gas in the wake will be at lower pressure (blue regions below the red line). Approaching the plane, the pressure increases overall. Once the cloud punches through the disk, the ambient pressure begins to drop, and the leading (faster) cloud gas (thin blue line below the red line beyond $240$~Myr) drops more rapidly than the reference value at the center of mass. 

 The dense, cold cloud gas of the initial conditions is nearly gone at the end of the burn-in phase (around $140$~Myr), leaving only warm gas at $T\sim10^4$~K. Since the temperatures in the three thermal regimes (hot, warm, cold) are mostly constant, the densities (Fig.~\ref{f:thermalevol}c) follow a shape similar to the pressure (Fig.~\ref{f:thermalevol}a). At around $100$~Myr, mass starts to flow from the hot component at $T=2.2\times 10^6$~K to the warm gas. Once the cloud approaches the plane, the pressure increases such that density and temperature can access the two thermal equilibrium solutions. The warm phase ($\sim10^4$~K) tries to transit to the cold phase, but dynamics prevent a full conversion, so that it reaches only the unstable neutral phase \citep{2018ApJS..238...14M}. The onset of the transition can be seen around $150$~Myr and $200$~Myr (Fig.~\ref{f:thermalevol}b), with gas moving from $T=10^4$~K to lower temperatures. While the pressure at $P/k_B>3\times 10^3$~K~cm$^{-3}$ should suffice for warm {\em and} cold neutral gas to co-exist at the same pressure (compare to Fig.~\ref{f:coolingfunc}), dynamical heating competes with radiative cooling \citep[][though in a different context]{2001ApJ...557L.121G}, suppressing the full formation of a cold phase.

 Most of the cloud's thermal state and thus the presence of neutral gas is determined by the ambient pressure \citep{1995ApJ...453..673W}, while dynamical effects lead to a gradual transition between the stable phases. Yet, conversion from warm to cold neutral medium is not complete by any means, rather, any gas leaving the warm neutral phase cools down only to the unstable neutral phase.  
  
 
 \subsection{Time Evolution and Inclination Angle}\label{ss:timeangle}
\begin{figure*}
   \includegraphics[width=\textwidth]{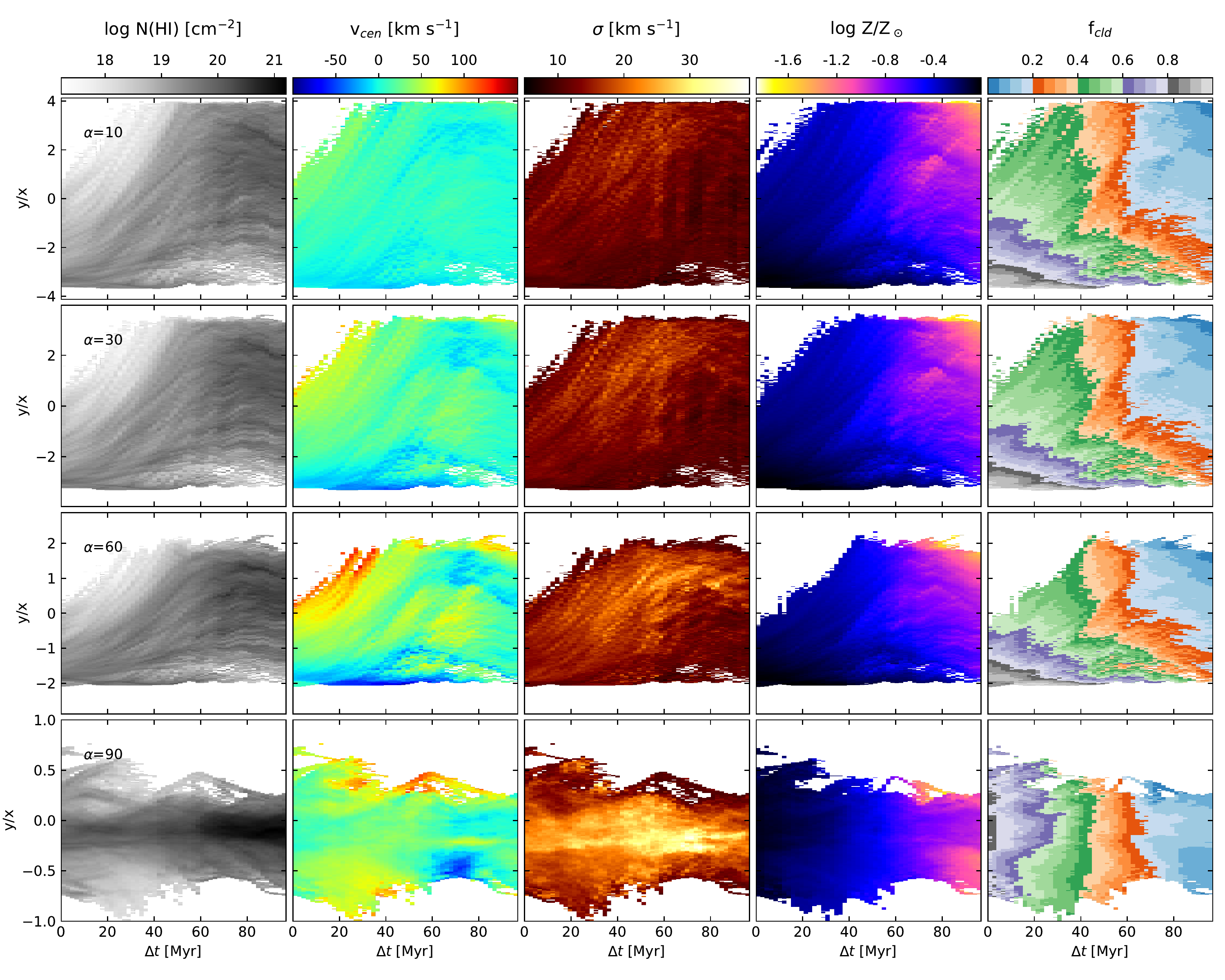}
   \caption{\label{f:allprof}Laterally averaged profiles along the cloud's long axis, against time after burn-in ($140$~Myr), for four inclination angles. Quantities shown and color schemes are identical to those in Fig.~\ref{f:allquant}. The ordinate shows the position along the cloud's long axis, in units of the simulation domain's lateral extent. From left to right: Column density, centroid velocity, non-thermal line-width, metallicity, and cloud mass fraction. Grey tones in the cloud mass fraction indicate highest fraction of original material. Therefore, material is accreted predominantly in the tail (top), and not in the head (bottom) of the cloud. All inclination angles except $\alpha_i=90^\circ$ (head-on) show a gradient in metallicity and cloud mass fraction.}
\end{figure*}

 Figure~\ref{f:allprof} shows the time evolution between $140$ and $240$~Myr (see Fig.~\ref{f:probtime}) of the cloud profiles in dependence of inclination angles, from $\alpha_i=10^\circ$ (nearly in the plane of sky) to $\alpha_i=90^\circ$ (cloud moving toward observer). Quantities and color schemes are identical to those used in Figure~\ref{f:allquant}.  
 
 The cloud stretches out and develops a head-tail structure, which then breaks up into fragments moving at different speeds as indicated by the blue-shifted material throughout the cloud for larger $\alpha_i$ (centroid velocities, 2nd column). Metallicities and cloud mass fractions drop with time, indicating contamination of the cloud by ambient material. Original cloud material is indicated by grey tones (cloud mass fraction, 5th column), while ambient gas is shown in blue tones. At a given time, the cloud mass fraction decreases from head (bottom) to tail (top) for all angles but $90^\circ$. Therefore, the tail is contaminated faster than the head, suggesting contamination via accretion into the cloud wake rather than by sweep-up of material by the head.  Despite the peloton effect and accretion, finding original cloud material near the leading part of the cloud is more likely than toward the trailing part, consistent with the results of \citet{2017ApJ...837...82H}. The cloud is predominantly contaminated by entraining  ambient material in its wake, and not via sweepup and compression. If sweepup and compression of ambient material were the main accretion mechanism, the leading part of the cloud should contain relatively {\em less} original cloud material than the trailing part. \citet{2017ApJ...842..102G} point out that for supersonic motions, condensation at the head of the cloud can be suppressed because a bow shock develops, but that material still can condense in the wake. At $\sim 300$~km~s$^{-1}$ (Fig.~\ref{f:trajectories}), the cloud travels supersonically with respect to the background medium. 
  
\begin{figure*}
   \includegraphics[width=\textwidth]{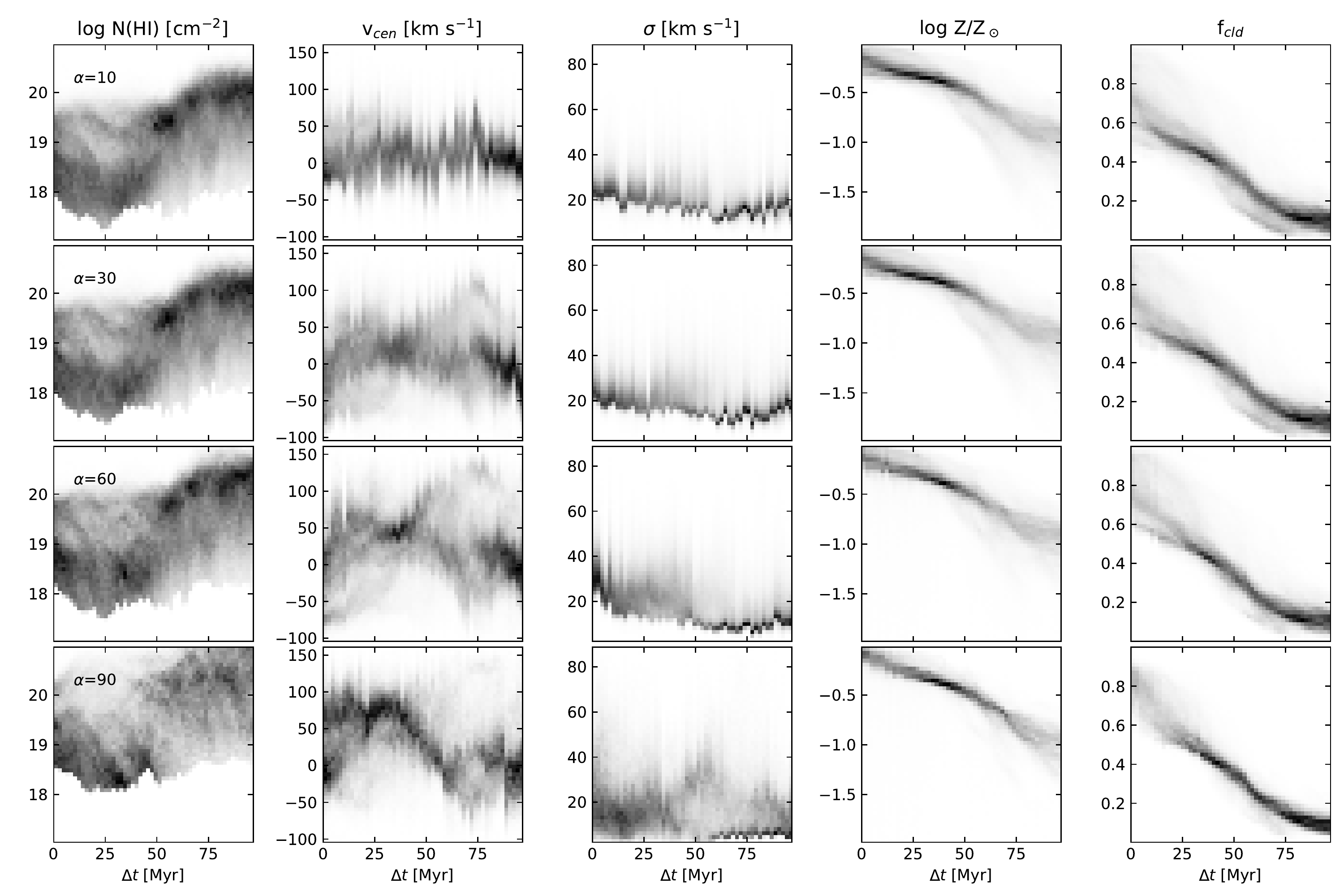}
   \caption{\label{f:allhist}Histograms of all quantities shown in Fig.~\ref{f:allprof}, against time after burn-in ($140$~Myr), for all inclination angles. From left to right: Column density, centroid velocity, non-thermal line-width, metallicity, and cloud mass fraction.}
\end{figure*}

Histograms of characteristic cloud measures (Figure~\ref{f:allhist}) highlight how accretion affects the overall cloud properties. The average column density increases with time, while the average metallicity and cloud mass fraction drop by more than a factor of $10$. The width of both metallicity and cloud mass fraction distributions decrease with time for larger inclination angles. The cloud accretes ambient material, replacing the original cloud material, and thus resulting in a more homogeneous composition. For $\alpha=10^\circ$, the metallicity distribution develops a tail toward lower values, resulting in a spread of over $1.5$~dex. For small inclination angles, more material at low column densities contributes to the lines-of-sight. Since low column densities are more likely to arise from accreted material (Sec.~\ref{ss:colmetal}), (for our model parameters) lower metallicities indicating ambient gas.

The cloud decelerates once it has punched through the disk. This reduces the drag along the cloud, and thus the velocity differences of accreted material between different cloud locations. Therefore, the overall velocity dispersion drops, and the cloud becomes dynamically more coherent.

Only extreme inclination angles ($\alpha_i=10^\circ,90^\circ$) affect the overall appearance of the cloud (Fig~\ref{f:allprof}). For small inclination angles, relative velocities along the cloud trajectory are suppressed, and therefore, the peloton effect cannot be observed. Metallicity estimates and distributions are consistent except for fully head-on trajectories ($\alpha_i=90^\circ$). Because of the large aspect ratio of cloud length over width, only $\alpha_i\approx90^\circ$ leads to confusion in metallicities. Variations in metallicities and cloud mass fraction along the cloud's long axis are consistent except for $\alpha_i\approx90^\circ$.

The probability to find original cloud material decreases from head to tail (Fig.~\ref{f:allprof} last column showing the cloud mass fraction), and it strongly depends on position, time and inclination angle. Note that finding original cloud material is related, but not identical to finding sight-lines with high original cloud material content. The difference can be seen by comparing the cloud mass fractions in Figs.~\ref{f:allquant} and Fig.~\ref{f:allprof}. Small pockets of sight-lines with $\sim50$\% cloud material survive, yet their area coverage is vanishingly small compared to the cloud area (See also Sec.~\ref{ss:colmetal}).

Distributions (Fig.~\ref{f:allhist}) are differently affected by inclination angles. Column densities generally increase for higher $\alpha_i$, but their distribution narrows, while the centroid velocities show the reverse effect, since for low $\alpha_i$, the lag of cloud material along the trajectory does not contribute to the velocity spread. At larger inclination angles, at least two centroid velocity components are discernible. Linewidths also increase with increasing $\alpha_i$, as more gas is mixed along the line-of-sight, leading to velocity crowding. Metallicities and cloud mass fraction distributions are mostly unaffected by the viewing angle, suggesting that the latter two -- especially metallicities in terms of observables -- are robust against variation in $\alpha_i$. 

\subsection{Velocity Lags and Metallicity}\label{ss:lagvelo}
\begin{figure*}
  \includegraphics[width=\textwidth]{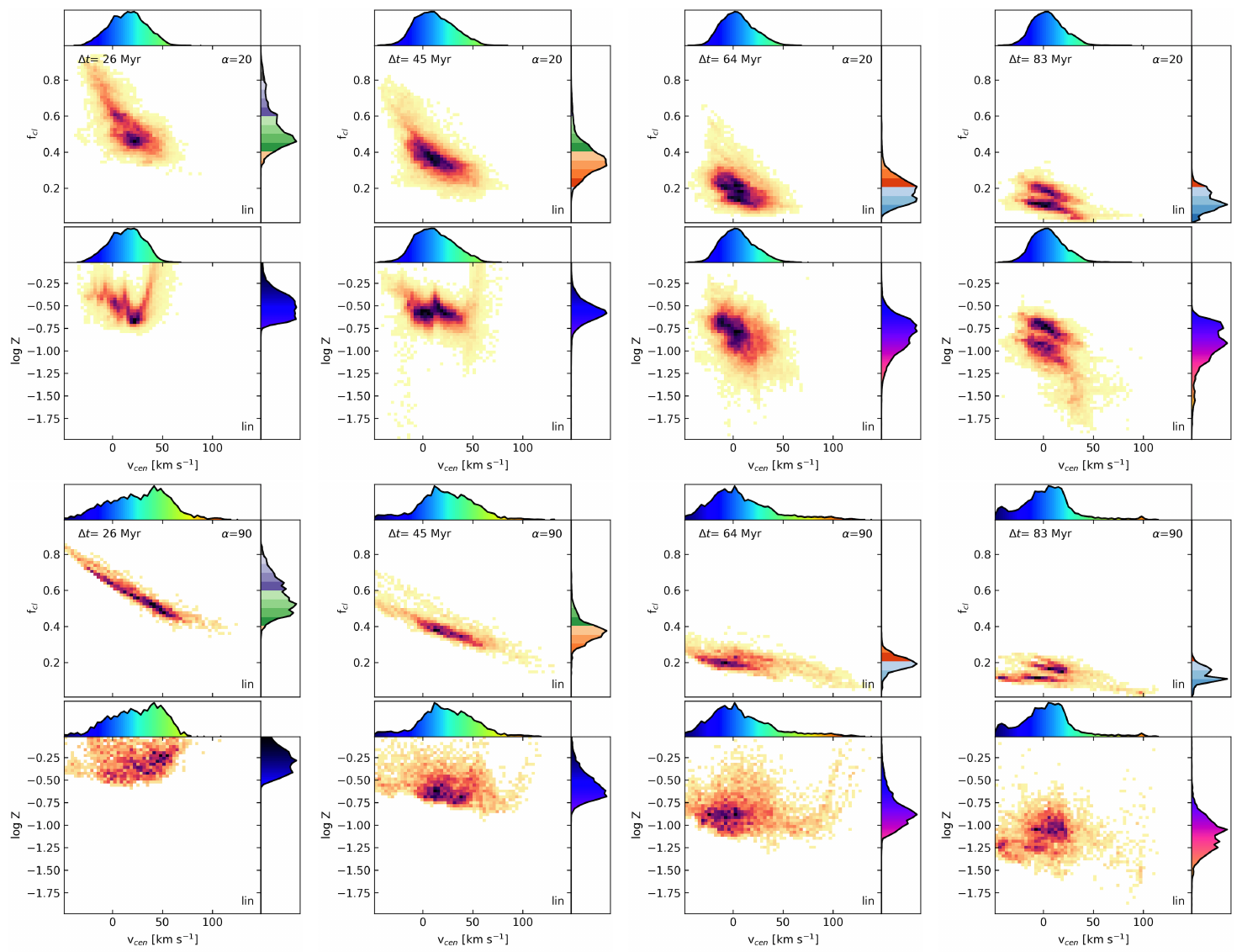}
  \caption{\label{f:lagmetal}Time history (columns) of the cloud mass fraction $f_c$ (top) and metallicity $\log Z$ (bottom) as function of the centroid velocity ($x$-axis), for several inclination angles (rows). Times $\Delta t$ are given after burn-in ($140$~Myr). Cloud mass fractions systematically drop with higher (red-shifted) velocities, and with increasing time. For the metallicities, both statements are generally true, though the viewing angles can hide the signature. Color scales of marginalized histograms are identical to those of Fig.~\ref{f:allquant}.}
\end{figure*}
 
Our discussion suggests that lower cloud mass fractions and ambient (in our case, lower) metallicities should be correlated with larger velocity lags along the cloud. Figure~\ref{f:lagmetal} confirms this expectation for early times. The 2D histograms of cloud mass fraction (top) against centroid velocity display a distinct negative gradient from small (${\rm v}_{\rm cen}=0$) to large lags (${\rm v}_{\rm cen}>0$).
Metallicities (bottom) mostly follow the same trend. The correlation becomes more pronounced for larger $\alpha_i$, since turbulent lateral motions of the cloud material do not contribute. At late times, both cloud mass fractions and metallicities develop two distinct peaks, both spread out over similar velocity ranges. This is a dynamic signature of the peloton effect: original and ambient material are mixed throughout the cloud, traveling at similar velocities. 
 
\subsection{Column Densities and Metallicities}\label{ss:colmetal}
\begin{figure*}
  \includegraphics[width=\textwidth]{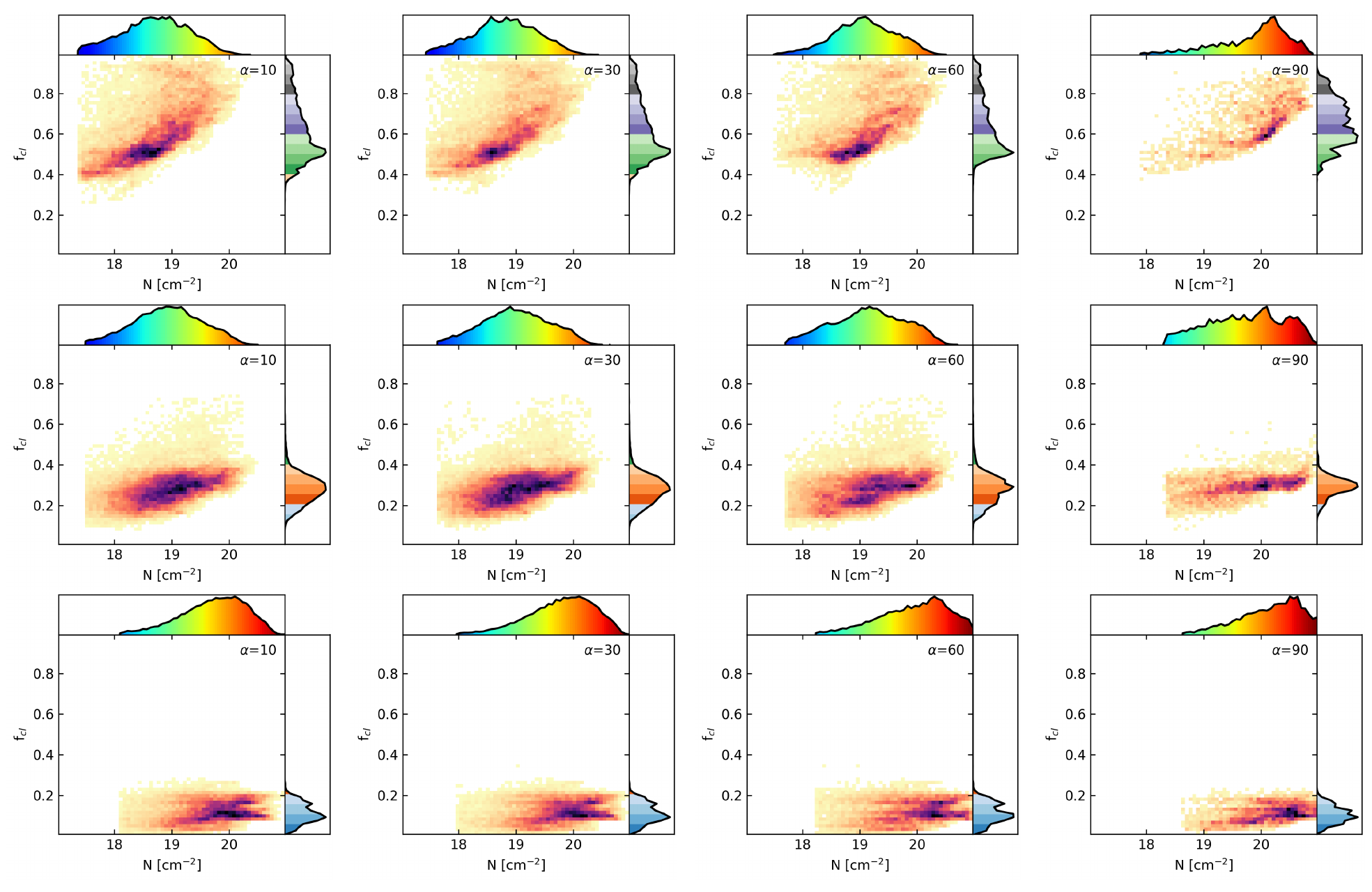}
  \caption{\label{f:colmetal}2D histograms of cloud mass fraction against column density, to test for correlation between metallicity and column density. A positive correlation indicates mixing between cloud and ambient gas. Colors stand for the number of sight lines. Times are $\Delta t=60, 80,100$~Myr (top to bottom). Different viewing angles are arranged horizontally, from in-the-plane (left) to a head-on view (right).}
\end{figure*}
Due to its proximity, the HVC complex with probably the highest number of abundance measurements is Complex C \citep{1999Natur.402..388W,2001ApJ...559..318R,2003AJ....125.3122T,2003ApJ...585..336C,2007ApJ...657..271C,2011ApJ...739..105S}. \citet{2007ApJ...657..271C} explore correlations between [O/H] abundances and H{\small{I}} column densities as a signature of mixing with ambient gas, hypothesizing that higher column densities should show less contamination with ambient material. While present, the observed trend is weak due to measurement uncertainties. The fourth column of Fig.~\ref{f:allquant} provides a modeler's version of Fig.~8 of \citet{2007ApJ...657..271C}. In the center row, the cloud mass fraction at the cloud edge (around $y=0$) is lower than that of the cloud's main body. These regions coincide with lower column densities (1st column). Note that while \citet{2007ApJ...657..271C} would expect an anti-correlation between metallicity and column density, for our model parameters a positive correlation between column density and metallicity indicates mixing between ambient and cloud gas. 

Figure~\ref{f:colmetal} provides a more detailed view, showing the joint distributions of cloud mass fraction as a proxy for metallicity, and column density. We show the cloud mass fraction, since the metallicities are model-dependent. A positive correlation between cloud mass fraction and column density indicates mixing and accretion. Indeed, for the first two time instances (top and center row), all viewing angles suggest a positive trend. When the cloud is about to punch through the disk, column densities rapidly increase, and a substantial amount of ambient material is accreted. This leads to a flat distribution with cloud mass fraction -- most of the neutral gas is now stemming from the ambient medium.

The specific choice of cloud and halo metallicities in our model precludes quantitative predictions of metallicities within any given cloud. Yet, the metallicity {\em gradient} (Fig.~\ref{f:allprof}) across a cloud provides an opportunity to distinguish between enrichment or dilution of cloud material, which will occur when the cloud's metallicity is lower or higher than that of the ambient gas, respectively. In the case of enrichment, the metallicity gradient along the cloud should be positive, since material of higher metallicity gets successively entrained in the cloud's wake (Fig.~\ref{f:masshistory}). In the case of dilution (as in our model cloud), the metallicity gradient should be negative, as low-metallicity gas gets entrained. While specific values will depend on several factors including the metallicity contrast between the cloud and the ambient gas, estimating the metallicity gradient across a given cloud even by two or three sight lines may provide information about the relative metallicities. 

The Smith Cloud can serve as an example. Although three metallicity measurements for the Smith Cloud \citep{2016ApJ...816L..11F} do not show a clear gradient, the two sight-lines close to the head of the cloud show higher values ([S/H]$=-0.09\pm0.33$ and $-0.14\pm0.13$), while the cloud tail has a lower value of [S/H]$=-0.58\pm0.20$. If we group the first two values together, the above reasoning would lead us to conclude that the Smith Cloud's material has been diluted (not enriched) by ambient gas of lower metallicity. On the other hand, \citet{2017ApJ...837...82H} 
focus on the two higher metallicity estimates and conclude that their models are consistent with 
the Smith Cloud having been enriched over time by high-metallicity ambient gas, but this does not explain
the lower metallicity measured in the cloud wake. More observational metallicity constraints, in both the cool clouds and the hot ambient gas, are needed for further progress.

Fig.~\ref{f:colmetal} provides a measure for the probability to find sight-lines with high original cloud material content, at a given point in the cloud's evolution. While the overall cloud mass fraction drops with time, pockets of sight-lines with a high cloud mass fraction survive (center row, lower inclination angles). Yet their area coverage is small compared to that of accreted material. For example the probability to find sight-lines with $f_{cl}>0.5$ at $\Delta t=80$~Myr is less than $1$\%.


\section{Discussion}\label{s:discussion}

We presented an analysis of a hydrodynamical simulation of a gas cloud traveling through the Galactic halo, determining the probability to find original cloud material within the observed cloud at any given time and viewing angle. The model assumes cloud formation via infall, possibly due to stripping from dwarf satellites. Other origin scenarios for cloud formation have been discussed elsewhere \citep[e.g.][for in-situ formation]{2018MNRAS.476..852J}.

\subsection{Caveats}\label{ss:caveats}
\subsubsection{Missing Physics}\label{sss:missingphysics}
We leave out the effects of thermal conduction and magnetic fields. \citet{2016ApJ...822...31B} investigated the role of heat conduction with 3D adaptive mesh refinement models, concluding that electron thermal conduction does play a role in the evaporation of clouds of low column densities. Though applied to a different scenario, namely the acceleration of cold clouds in a galactic wind, the underlying physics is similar. Two-dimensional models of cloud disruption including radiative cooling and heat conduction \citep{2007A&A...472..141V,2007A&A...475..251V,2017MNRAS.470..114A} have shown that heat conduction can help to stabilize the cloud against disruption. The latter authors exploit the advantages of reduced dimensions and explore a range of cloud radii and masses. \citet{2021MNRAS.502.1263K} demonstrate numerically that magnetic fields might suppress thermal conduction \citep{1962pfig.book.....S} and thus help cool gas "survive". Yet they point out that thermal conduction -- even if suppressed -- still affects the evolution of a cloud.

In a similar study of cloud acceleration in hot winds, \citet{2015MNRAS.449....2M} conclude that, when using tangled magnetic fields, the clouds start to move with the background medium in near pressure equilibrium, rather than being disrupted. The magnetic field suppresses the dynamical instabilities \citep[see also][]{2017ApJ...845...69G} while it links the cloud to the ambient gas via sweep-up of field lines, thus increasing the drag force above hydrodynamic values. To capture the effect of magnetic fields fully, three-dimensional simulations are necessary \citep{2018ApJ...865...64G}, since in two dimensions, interchange modes cannot develop \citep{2007ApJ...671.1726S}. Still, magnetic fields seem capable of suppressing some of the condensation of ambient material into the wake \citep[][see also \citealp{1965ApJ...142..531F} for the suppression of thermal instability in the presence of magnetic fields]{2018ApJ...865...64G}, therefore we expect that including magnetic fields in our models would slow down the cloud and suppress cloud disruption, resulting in less mass lost and therefore in a higher probability to detect cloud material along the tail (Fig.~\ref{f:probtime}). 

\subsubsection{Metallicities and Disk Structure}
Our choice of initial cloud and specifically halo metallicities will reduce the condensation of halo gas in the wake of the cloud compared to more realistic (higher) halo metallicities. Therefore, our contamination estimates are {\em lower} limits, since higher halo metallicities would lead to more gas condensing in the wake of the cloud via cooling. Also, metallicity gradients would be less steep than found in our analysis. Omitting the thin disk with its colder and denser (neutral) gas component in our simplified halo model also suppresses accretion of ambient gas.

\subsubsection{Beam Smearing}
 We assume pencil beams for both absorption and emission. All HVC metallicity estimates are derived by combining H{\small{I}}~21~cm measurements taken at a finite beam width with UV metal-line measurements taken at an infinitesimal beam (pencil) width and are therefore sensitive to beam-smearing effects, in which small-scale (sub-beam) structure in the radio data could impact the actual H{\small{I}} column along the UV pencil beam. \citet{1999ASPC..166..302W,2001ApJS..136..537W} quantified this effect for a few lines-of-sight, extending to a list of H{\small{I}}~$21$~cm columns for all available HVC sight lines \citep{2011ApJ...728..159W}. They estimated that, typically, H{\small{I}} columns are accurate within $10-25$\%, or $\pm 0.04-0.10$ in $\log N_{\mathrm{H}}$. We will leave the consideration of beam-smearing effects for a future contribution.
 
 \subsubsection{Impact Velocity}\label{sss:impactvel}
 
 The cloud's velocity relative to the ambient gas is not only relevant for hydrodynamical considerations since instabilities can be suppressed for supersonic motions \citep{1961hhs..book.....C} and therefore the accretion mechanism changes \citep{2017ApJ...842..102G}, but it also could serve as a check against observations. Yet, three-dimensional velocity information for HVCs is available only under rare circumstances. 
 
 \citet{2008ApJ...679L..21L}  determine the three spatial velocity components of the Smith Cloud, with a total space velocity of $V_{tot}=300$~km~s$^{-1}$ and a vertical component of $73\pm 26$~km~s$^{-1}$ at which the cloud is approaching the disk. Its orbit is highly inclined and prograde, so that the Smith cloud's  velocity relative to the ambient gas would expected to be much less than $V_{tot}$.  \citet{2016MNRAS.462L..46H} proposed a method to reconstruct the three-dimensional cloud propagation direction required to calculate $V_{tot}$, based on the cloud morphology in the position-velocity plane and $v_{LSR}$. Applying the method to a subset of clouds identified in HIPASS \citep{2002AJ....123..873P}, they estimate $V_{tot}$ of up to $>300$~km~s$^{-1}$ for a few clouds. Yet, these estimates, and the resulting relative speed with respect to the ambient gas, strongly depend on the assumed halo gas rotation model.
 
 Theoretical estimates of the infall velocity depend strongly on the cloud's starting point, on the gravitational potential, and on whether drag is included. \citet{2020ApJ...903..101S} find infall velocities of $\sim350$~km~s$^{-1}$ without drag and a logarithmic potential (Fig.~\ref{f:trajectories}), while \citet{1997ApJ...481..764B} including drag estimate terminal velocities of $~\sim 150$~km~s$^{-1}$. Self-consistent models in which the cloud is dropped in a stratified halo \citep{2009ApJ...698.1485H,2021MNRAS.501.5330S} reach cloud velocities between $150$ and $>300$~km~s${-1}$. Wind-tunnel experiments explore ranges of $100$--$350$~km~s$^{-1}$ \citep[e.g.][]{2011ApJ...739...30K}.

\subsection{Consequences for Observations}\label{ss:consequences}
\noindent(1) The peloton effect (Sec.~\ref{sss:peloton}) could be detected in three ways: (a)  High-resolution (interferometry) observations at the leading head of a cloud would exhibit "unordered" velocity fields not following a systematic velocity gradient from head to tail (Fig.~\ref{f:allprof}). (b) The (thermally) cooler head would show a large non-thermal component. This could be looked for in H{\small{I}}~$21$~cm data as well as in UV absorption lines (Fig.~\ref{f:allprof}). (c) Joint distributions of metallicity and centroid velocities would show double or multiple peaks (Fig.~\ref{f:lagmetal}). \\

\noindent(2) Using metallicities as diagnostics for cloud origin is complicated mostly by substantial contamination especially of the trailing cloud parts due to accretion of ambient gas (Figs.~\ref{f:probtime}, \ref{f:masshistory}, \ref{f:allprof}, see also \citealp{2017ApJ...837...82H}. Though the leading cloud part is less prone to accrete material specifically at supersonic motions, the peloton effect will eventually also lead to contamination, flattening the metallicity gradient to a point where distinguishing between original and accreted gas might be impossible.\\

\noindent(3)
Not only does the mass fraction of original cloud material drop with time, but also the area coverage of sight-lines with high original cloud mass fraction decreases rapidly. The probability to find such sight-lines strongly depends on inclination angle and time (Figs.~\ref{f:allquant}) and \ref{f:allprof}). Fig.~\ref{f:colmetal} suggests that the probability to find sight-lines with $f_{cl}>0.5$ is vanishingly small for evolved clouds. Therefore, in all likelihood, metallicity measurements in HVCs will not represent original cloud properties, but metallicity gradients will provide insight about cloud contamination.\\ 

\noindent(4) While the H{\small{I}} column densities vary over nearly two orders of magnitude at any time, the metallicities are spread over half a dex at most at any given time (Fig.~\ref{f:allhist}). Metallicities drop with time due to accretion well beyond observational uncertainties \citep{2016ApJ...816L..11F}, yet an individual snapshot shows a largely uniform metallicity. The spread is larger than observational uncertainties at any given time, allowing observational distinction between cloud and ambient material only {\em if their metallicities differ sufficiently}.  Viewing angles do not affect the metallicity estimates.\\

\noindent(5) If the ambient metallicity is known, variations of the metallicity along the cloud can provide information about whether gas has been mostly accreted via compression at the leading head or via condensation in the trailing tail.\\

\noindent(6) Metallicity gradients along the cloud trajectory can provide insight about the relative metallicities of cloud and ambient gas, based on the observation that ambient material is successively entrained in the cloud's wake
(Fig.~\ref{f:allprof}, Sec.~\ref{ss:colmetal}).\\

\noindent(7) Correlations between metallicity and column density can indicate the presence of cloud gas mixing with ambient gas \citep{2007ApJ...657..271C}. Figure~\ref{f:colmetal} suggests that such a correlation would strongly depend on the evolutionary state of the cloud -- if such a concept is applicable at all. A "pristine" cloud traveling through the halo would show a correlation between metallicity and column density, but once the majority of a cloud's mass is actually coming from the ambient gas, the correlation between metallicity and column density should be flat.

\subsection{Outlook}
While our parameter choices do allow us to set strong lower limits on contamination, ultimately, exploring a range of halo and cloud metallicities, cloud masses, and trajectories might seem advisable to further constrain how accretion affects realistic clouds. Yet, based on our parameter choices, we do not expect results to change qualitatively.

Our model addresses the  evolution and contamination of an infalling cloud, but similar contamination mechanisms might be expected for outflowing clouds driven by Galactic winds \citep[e.g.][]{2018ApJ...855...33D,2020ApJ...888...51L}. Depending on the balance between ram pressure acceleration by the wind and gravitational acceleration, corresponding models might be closer to wind-tunnel experiments than envisaged here.

\section{Conclusions}
We use hydrodynamic simulations of infalling high-velocity clouds run with a modified version of Athena \citep{2008ApJS..178..137S} to probe how reliably metallicity measurements can identify 'original' cloud material over time, along the cloud, and under a range of viewing angles. The model parameters are chosen to provide conservative (strong) limits on cloud contamination by ambient material. Our analysis has led us to conclude:\\

(1) The cloud is contaminated nearly linearly with time, until most of the original cloud material has been replaced by accreted gas (Fig.~\ref{f:allhist}). The cloud is well defined at any {\em fixed} time, but not {\em across} time, and thus  is not defined by constancy in its constituent gas particles, but by self-propagating density and pressure perturbations. The energy necessary to reform the cloud is provided by the potential energy of the over-densities.\\
(2) Our super-sonically moving cloud is to some extent contaminated at its leading edge by sweep-up, but mostly along the cloud tail, due to thermal condensation into the wake of the cloud (Fig.~\ref{f:allprof}). The thermal state of cloud gas - whether original or accreted - is largely controlled by the ambient pressure (Fig.~\ref{f:thermalevol}) and the cloud dynamics.\\
(3) The cloud dynamics can get extremely complex, with initially lagging material that is shielded from ram pressure catching up to the head, and even passing the head, of the cloud (peloton effect). This would affect centroid velocity gradients and thus drag estimates \citep{2007ApJ...656..907P}, and it could also introduce uncertainties in metallicity gradients.\\
(4) While the dynamical quantities (centroid velocity and velocity dispersion) depend on the inclination angle, the overall metallicities are unaffected by inclination within observational uncertainties, due to identification of individual velocity components (Fig.~\ref{f:allhist}). Only for small inclination angles (i.e. clouds mostly in the plane of sky), accreted material at low column densities will display metallicities (given our model parameters) lower by more than $1$~dex compared to the bulk of the cloud.\\
(5) The correlation between metallicities and column densities along a range of sight lines not only indicates mixing \citep{2007ApJ...657..271C}, but also could indicate the evolutionary stage of the cloud.\\

\appendix

\section{Modifications to Athena}\label{s:athenamod}
The code is available at FH's {\tt github} repository. The initialization file can be found under {\tt src/prob/fheitsch/hvc.c}.


\subsection{PPV Cubes}\label{ss:ppvcubes}
Because of the high resolution requirements, we implemented an on-the-fly analysis, including the calculation of position-position-velocity (PPV) cubes for selected viewing angles. All lines-of-sight can be calculated at initialization of the simulation. The block decomposition used in the parallelization requires that each processor dumps its version of a rotated PPV cube. These are then added and stitched together in a post-processing step. This approach limits the radiative transfer to optically thin emission and/or background absorption. Self-absorption cannot be modeled. We calculate spectra for two species.

The H{\small{I}}-$21$~cm emission is determined for each velocity channel ${\rm v}$ as the integral along the line of sight $s$,
\begin{equation}
  T_b({\rm v}) = c_{\mathrm{H{\small{I}}}}^{-1} \int n(s)\phi({\rm v}-{\rm v}(s))\,ds,
\end{equation} 
where $n(s)$ is the local density in cm$^{-3}$ along the line-of-sight for gas with $T<10^4$~K, and the function 
\begin{equation}
  \phi({\rm v}-{\rm v}(s)) = \frac{1}{\sqrt{\pi}\Delta v_{\mathrm{th}}(s)}\exp\left(-\left(\frac{{\rm v}-{\rm v}(s)}{\Delta {\rm v}_\mathrm{th}(s)}\right)^2\right)
\end{equation}
describes the thermal broadening with the local thermal width $\Delta {\rm v}_{\mathrm{th}}(s)=\sqrt{2k_BT(s)/m_{\rm H}}$, where $m_{\rm H}$ is the hydrogen atom mass, and $k_B$ the Boltzmann constant. The constant  $c_{\mathrm{H{\small{I}}}}=1.823 \times 10^{18}$ \,cm$^{-2}$\,(K km s$^{-1}$)$^{-1}$. Each channel is $\Delta {\rm v}=1$~km~s$^{-1}$ wide.
The brightness temperature can be converted into a total column density via
\begin{equation}
  N(\mathrm{H\small{I}}) = c_{\mathrm{H{\small{I}}}} \,\int T_b({\rm v})dv. \label{e:Nemiss}
\end{equation}

The absorption by a tracer species (for example singly ionized sulfur, S{\small{II}}, 
\citealp{2016ApJ...816L..11F}) is calculated based on the local metallicity,
\begin{equation}
  Z = Z_\mathrm{c}C_\mathrm{c} + Z_\mathrm{h}C_\mathrm{h},\label{e:metallicities}
\end{equation}
where the subscripts $c/h$ denote cloud/halo material, and with the sum of the color fields identifying cloud and halo material, $C_\mathrm{c}+C_\mathrm{h}=1$. The color field for the cloud $C_\mathrm{c}$ is defined via equation~\ref{e:defcloud}. We write the absorption coefficient as
\begin{equation}
  \kappa(s) = \sigma_{\mathrm{S{\small{II}}}}Z(s)n(s),
\end{equation}
assuming $\sigma_{\mathrm{S{\small{II}}}}=10^{-22}$~cm$^2$ as a cross section for the tracer. 
The exact choice is irrelevant for our purposes, and is solely motivated by rendering the resulting absorption line as optically thin, i.e. the optical depth at a given velocity channel $v$ is given by
\begin{equation}
  \tau({\rm v}) = \int\kappa(s)\phi({\rm v}-{\rm v}(s))\,ds.\label{e:Nabs}
\end{equation}
The column density per channel $N({\rm v})$ can then be calculated as $\tau({\rm v})/\sigma_{\mathrm{S{\small{I}}}}$. We note that we use the same mean molecular weight for H{\small{I}} and the tracer species. Further, we neglect any issues arising from varying ionization degrees \citep{2016ApJ...816L..11F}. In that sense, our 'metal absorber' is just a generic quantity, providing the simplest possible model. The closest equivalent to observational measurements would be metallicity estimates via O{\small{I}}/H{\small{I}} coupled via charge exchange \citep{2003ApJ...585..336C}.


\subsection{Internal Energy}
The Athena stock version applies density and pressure floors after the reconstruction step as a safeguard to prevent negative
densities and/or pressures in extreme flow situations. For problems involving radiative losses, the floor values can lead
to unreasonably high heating rates, leading to point-wise "explosions". Therefore, we implemented an internal energy integration following \citet{2014ApJS..211...19B}. Details are discussed by \citet{2016MNRAS.462.2777G} and \citet{2019MNRAS.489...52F}.


\subsection{Time Stepping}\label{ss:timestepping}
To increase the stability of the code in the presence of strong temperature gradients and high-Mach number shocks, we implemented a time-variation-diminishing third-order Runge-Kutta integrator \citep{1998MaCom..67...73G} in Athena. While slower than the CTU and VL integrators of the Athena 4.2 stock version, the benefit of improved stability while being able to run at Courant numbers of $0.5$ seemed a reasonable trade-off. The RK3 architecture has the advantage that arbitrary source terms can be included with ease at third order in time, i.e.
at the same order as the hydrodynamical fluxes. This is especially of advantage in the case of energy source terms such as cooling (App.~\ref{s:thermalphysics}) or co-moving grids (App.~\ref{s:comovinggrid}). Despite the higher time-accuracy, extreme flow situations, specifically in the presence of radiative losses, can occasionally lead to negative densities. In such cases, the integrator repeats the failed step at half the Courant timestep for the whole grid.


\subsection{Thermal Physics}\label{s:thermalphysics}
\begin{figure*}
  \includegraphics[width=\textwidth]{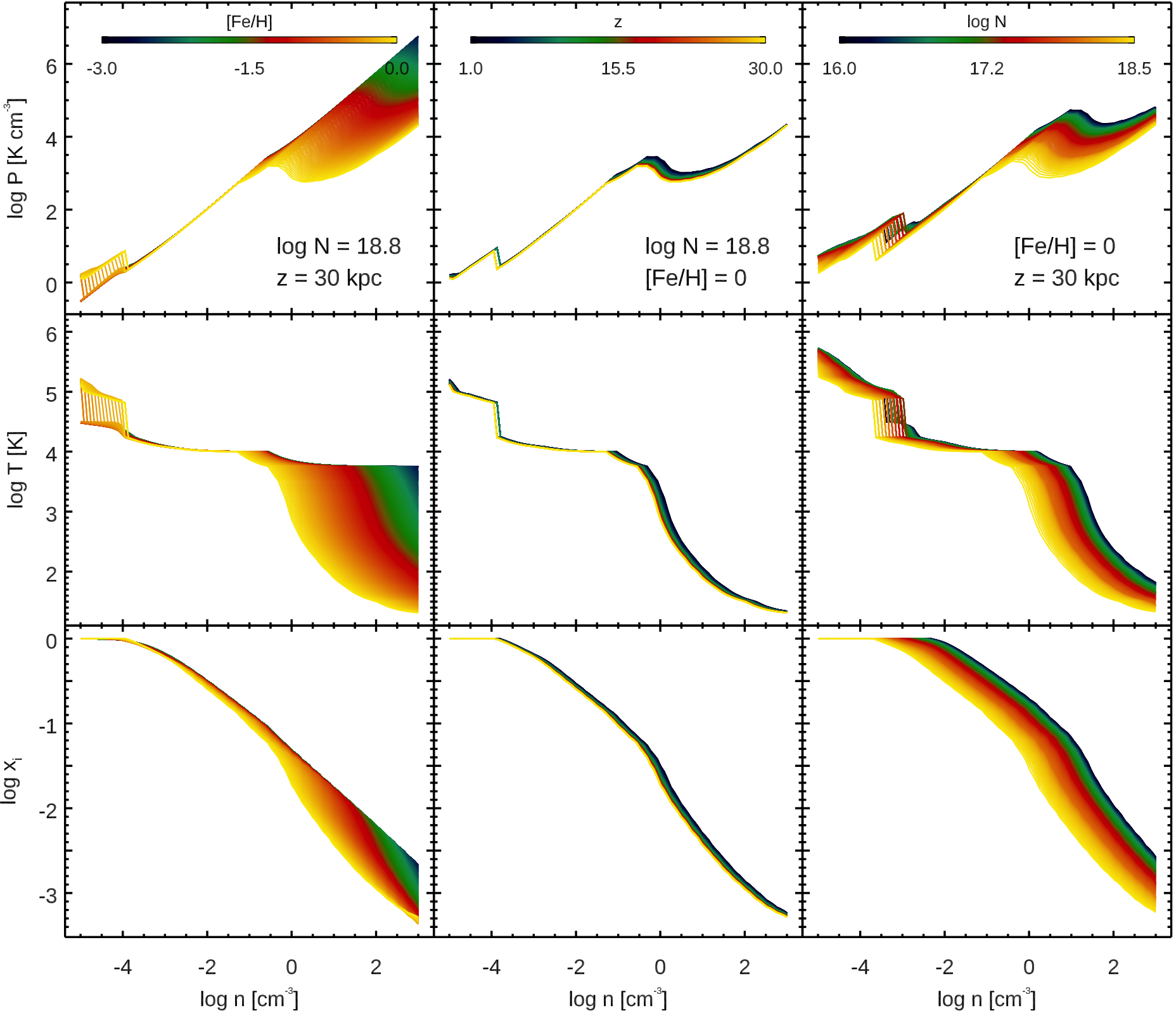}
  \caption{\label{f:coolingfunc}Thermal equilibrium pressure (top), temperature (center), and ionization fraction (bottom) against density, in dependence of metallicity (left), distance from plane (center), and shielding column (right). Metallicity and shielding column have the strongest effect on the thermal behavior of the gas. The ionization fraction is used to calculate the amount of neutral H{\small{I}}.}
\end{figure*}

Heating and radiative losses are implemented as a look-up table in five variables, namely the gas density $n$, the temperature $T$, the distance from the Galactic plane $z$ to account for reduction of soft UV radiation density, the column density $N$ to account for UV and x-ray shielding, and the metallicity $Z$. Tables were generated with Mappings-V \citep{1993ApJS...88..253S}. The $z$-dependence of the radiation field follows \citet{1995ApJ...453..673W}. Figure~\ref{f:coolingfunc} shows the thermal equilibrium pressures $P$, temperatures $T$, and ionization degrees against gas density $n$, in dependence of metallicity (left), distance above the plane (center), and shielding column (right). Metallicity and shielding can strongly affect the thermal behavior of the gas \citep{1995ApJ...453..673W}. The calculation of the ionization degree assumes collisional ionization equilibrium, since we are interested only in the overall dynamics rather than the details of the turbulent mixing layers or the ionization stages of various ions \citep{2010ApJ...719..523K,2010ApJ...718.1315G}. We use the ionization degree to determine the fraction of neutral gas during the analysis step. Note that a detailed treatment of the thermal physics across 'phase transitions' \citep[e.g.][]{2009ApJ...696..233S} including non-equilibrium cooling can change the nature of the transition and thus could eventually affect the efficiency of ambient gas condensation in the cloud wake \citep{2017ApJS..228...11G}.

Heating and cooling are implemented as sources $\dot{E}_{cool}$ to the total energy equation,
\begin{equation}
  \partial_t E + \nabla\cdot\left(\pmbu(E+P)\right) = \dot{E}_{cool}.
\end{equation}
To prevent over-shoots once thermal equilibrium is reached, and to avoid negative temperatures, we sub-cycle on the cooling using an adaptive stepsize Runge-Kutta-Fehlberg integrator, solving
\begin{equation}
 \frac{dT}{dt} = f(n,T,N,Z,z).\label{e:dTdt}
\end{equation}
The RKF integrator
subcycles on eq.~(\ref{e:dTdt}) until a full Courant timestep $\Delta t$ is reached. The resulting energy difference $\Delta E$
is used to calculate
\begin{equation}
  \dot{E}_{cool} = \frac{\Delta E}{\Delta t}.
\end{equation}


\subsection{Comoving Grid}\label{s:comovinggrid}
There are two advantages to have the simulation domain track the cloud. First, we can limit the simulation domain to a box around the cloud. This allows us to track the cloud for an arbitrary amount of time, as long as we provide the correct (time- and position-dependent) boundary conditions. \citet{2014ApJ...795...99G} and  \citet{2016ApJ...816L..18G} decided to solve this problem by using adaptive mesh refinement in a domain spanning the whole trajectory of the cloud,  refining only on the position of the cloud. Yet, since we are mostly interested in turbulent mixing, we decided to insist on a uniform spatial resolution. \citet{2008ApJ...680..336S} point out the second advantage, namely that a comoving grid can be realized by subtracting the center-of-mass velocity from the box, thus reducing the effect of truncation errors in the cloud dynamics. We follow their approach, adjusting the subtracted velocity such that the leading cloud edge is kept at a (nearly) constant position within the box \citep[see also][]{2017MNRAS.468.3184G}. The comoving grid is implemented as a change in position and velocity applied to the grid coordinates, updated together with the fluid variables. Thus, the grid positions are updated together with the RK3 integration. The RK3 integrator takes care of resetting the coordinates also, in case a timestep needs to be repeated (see Appendix~\ref{ss:timestepping}).


\section{ROHSA}\label{s:ROHSA}

Fitting the spectra to determine column densities and thus abundances met with two challenges. First, a single position-position-velocity cube can contain thousands of usable lines-of-sight, since we are not limited to background sources for absorption spectra. Second, a single line-of-sight may contain multiple velocity components, the number of which is not known in advance. The first challenge mandates a fast and efficient fitting mechanism, and the second challenge requires the fitting procedure to allow for an arbitrary number of Gaussian components, yet ideally without over-fitting the data. A first attempt to use Gaussian Mixture Models \citep{Reynolds2015} turned out to be unviable because of time requirements. We therefore decided to use a modified version of {\texttt{ROHSA}} \citep{2019A&A...626A.101M} that decomposes jointly the emission and absorption position-position-velocity cubes described in Appendix~\ref{ss:ppvcubes} into multiple velocity components using a Gaussian model.

\subsection{Model}\label{subsec:model-ROHSA}

The data are the measured brightness temperature $T_b(v,\rb)$ and optical depth $\tau(v,\rb)$ at a given projected velocity $v$. The proposed model
$\tilde T_b\big(v, \thetab^{em}(\rb)\big)$ and $\tilde \tau\big(v, \thetab^{\tau}(\rb)\big)$ are a sum of $N$ Gaussian $G\big(v, \thetab_n(\rb)\big)$
\begin{equation}
  \tilde T_b\big(v, \thetab^{em}(\rb)\big) = \sum_{n=1}^{N} G\big(v, \thetab^{em}_n(\rb)\big),
\end{equation}
\begin{equation}
  \tilde \tau\big(v, \thetab^{tau}(\rb)\big) = \sum_{n=1}^{N} G\big(v, \thetab^{\tau}_n(\rb)\big),
\end{equation}
with $\thetab^{em}(\rb) = \big(\thetab^{em}_1(\rb), \dots, \thetab^{em}_n(\rb)\big)$, $\thetab^{\tau}(\rb) = \big(\thetab^{\tau}_1(\rb), \dots, \thetab^{\tau}_n(\rb)\big)$, and where
\begin{equation}
  G\big(v, \thetab_n(\rb)\big) = \ab_n(\rb) \exp
  \left( - \frac{\big(v - \mub_n(\rb)\big)^2}{2 \sigmab_n(\rb)^2} \right)
\end{equation}
is parametrized by $\thetab_n = \big(\ab_n, \mub_n, \sigmab_n\big)$ with
$\ab_{n} \geq \bm{0}$ being the amplitude, $\mub_{n}$ the position, and
$\sigmab_{n}$ the standard deviation 2D maps of the $n$-th Gaussian
profile. The residuals are
\begin{equation}
  L^{em}\big(v, \thetab^{em}(\rb)\big) = \tilde T_b\big(v, \thetab^{em}(\rb)\big) - T_b(v,\rb),
\end{equation}
and 
\begin{equation}
  L^{\tau}\big(v, \thetab^{\tau}(\rb)\big) = \tilde \tau\big(v, \thetab^{\tau}(\rb)\big) - \tau(v,\rb).
\end{equation}

The estimated 
parameters $\hat{\thetab}^{em}$ and $\hat{\thetab}^{\tau}$ are defined as the minimizer of a cost function that includes the sum of the squares of the residuals,
\begin{equation}
  Q(\thetab^{em},\thetab^{\tau}) = \frac{1}{2} \, \lambda_{em} \, \norm{L^{em}\big(v, \thetab^{em}\big)}_2^2 + \frac{1}{2} \, \lambda_{\tau} \, \norm{L^{tau}\big(v, \thetab^{\tau}\big)}_2^2,
  \label{eq:2}
\end{equation}
where, $\lambda_{em}$, and $\lambda_{\tau}$ are hyper-parameters than tune the balance between the emission and absorption terms. Although the two terms in Eq.~\ref{eq:2} are combined in a single cost function, the parameters $\thetab^{em}$ and $\thetab^{\tau}$ are independent and the estimated parameters obtained using any arbitrary optimization algorithm would lead to the same result as if they were evaluated separately. However, such a combination allows us to add regularization terms as priors on the model parameters $\thetab^{em}$ and $\thetab^{\tau}$. 

Following \cite{2019A&A...626A.101M}, each parameter map of $\thetab^{em}(\rb)$ and $\thetab^{\tau}(\rb)$ is filtered using a Laplacian kernel to ensure that the solution is spatial coherent. In other words, adjacent pixels are forced to have a similar Gaussian decomposition. The following regularization term, itself containing energy terms, is added to 
the cost function given in Eq.~\eqref{eq:2}
\begin{align}
    R(\thetab^{em}, \thetab^{\tau}) &=
    \frac{1}{2} \, \sum_{n=1}^N \lambda^{em}_{\ab} 
  \|\Db\ab^{em}_n\|_2^2 + \lambda^{em}_{\mub} \|\Db\mub^{em}_n\|_2^2 \\
  &+
  \lambda^{\tau}_{\sigmab} \|\Db\sigmab^{\tau}_n\|_2^2 \\
  &+ \frac{1}{2} \, \sum_{n=1}^N \lambda^{\tau}_{\ab} 
  \|\Db\ab^{\tau}_n\|_2^2 + \lambda^{\tau}_{\mub} \|\Db\mub^{\tau}_n\|_2^2 \\
  &+
  \lambda^{\tau}_{\sigmab} \|\Db\sigmab^{\tau}_n\|_2^2,
\end{align}
where $\Db$ is the matrix perform the 2D convolution (see \cite{2019A&A...626A.101M} for complementary details), and $\lambda^{em}_{\ab}$, $\lambda^{em}_{\mub}$, $\lambda^{em}_{\sigmab}$, $\lambda^{\tau}_{\ab}$, $\lambda^{\tau}_{\mub}$, $\lambda^{\tau}_{\sigmab}$ are hyper-parameters than tune the balance between the different terms.

In addition, we also add the following regularization term to 
the cost function given in Eq.~\eqref{eq:2}
\begin{equation}
    R'(\thetab^{em}, \thetab^{\tau}) =  \frac{1}{2} \, \sum_{n=1}^N \lambda_{\mu} \norm{\mub^{em}_n/\mub^{\tau}_n - 1}_2^2 + \lambda_{\sigma} \norm{\sigmab^{em}_n/\sigmab^{\tau}_n - 1}_2^2,
    \label{eq:correlation}
\end{equation}
where $\lambda_{\mu}$, and $\lambda_{\sigma}$ are hyper-parameters than tune the balance between the different terms. These energy terms aim to maximize the correlation of the velocity fields and dispersion velocity fields between emission and absorption. In other words, these energy terms ensure that each Gaussian pair describing a component seen in absorption and emission has a similar velocity and a similar velocity dispersion. 

The full cost function is then
\begin{equation}
    J(\thetab^{em}, \thetab^{\tau}) = Q(\thetab^{em}, \thetab^{\tau}) + R(\thetab^{em}, \thetab^{\tau}) + R'(\thetab^{em}, \thetab^{\tau}),
    \label{eq:full-cost}
\end{equation}
and the minimizer is
\begin{equation}
    [\hat \thetab^{em}, \hat \thetab^{\tau}] = \underset{\thetab^{em},\thetab^{\tau}}{\text{argmin}}\ J(\thetab^{em},\thetab^{\tau}),
    \label{eq:6}
\end{equation}
wrt. $\ab^{em}_n \geq 0, \ab^{\tau}_n \geq 0\ \forall \, n \in [1, \dots, N]$. Following \cite{2019A&A...626A.101M}, this optimization relies on L-BFGS-B  (for Limited-memory Broyden–Fletcher–Goldfarb–Shanno with Bounds), a quasi-Newton iterative algorithm described by \cite{zhu_algorithm_1997} which allows for the positivity constraints of the amplitudes to be taken into account. 

The initialization of the optimization is carried out in two steps. First, the optical depth model $\thetab^{\tau}$ is adjusted using the second term in Eq.~\ref{eq:2}. For this step, we use the multi-resolution process from coarse to fine grid described in \citet{2019A&A...626A.101M}. Then the solution $\thetab^{\tau}$ is used to initialize $\thetab^{em}$. Finally, the full cost function given in Eq.~\ref{eq:full-cost} is used to perform the final optimization and to update $\thetab^{\tau}$ and $\thetab^{em}$, ensuring a spatially coherent solution with a correlated velocity field and velocity dispersion field for each Gaussian component. 

The decomposition is performed using {\tt ROHSA} hyper-parameters $N=6$, and $\lambda_{em}=\lambda_{\tau}=\lambda_{\ab}^{em}=\lambda_{\mub}^{em}=\lambda_{\sigmab}^{\tau}=\lambda_{\ab}^{\tau}=\lambda_{\mub}^{\tau}=\lambda_{\sigmab}^{\tau}=1$. The number of Gaussian $N$ is chosen to ensure that the signal is fully encoded. Note that due to regularisation, this does not imply that six Gaussians are used along each line of sight \citep{2019A&A...626A.101M}.

\subsection{Estimating physical parameters}\label{subsec:estimating-parameters}
The spatially coherent parameters $\hat \thetab^{em}, \hat \thetab^{\tau}$ obtained with {\tt ROHSA} allows us to extract the mean physical properties of the HVC gas along each line of sight.

The column density map is
\begin{equation}
    N(\mathrm{H\small{I}})(\rb) = C \, \ab_n^{em}(\rb) \sigmab_n^{em}(\rb).
\end{equation}
All the following quantities (metallicity, centroid velocity, and total velocity dispersion, see also eqs.~[\ref{e:vcen}] and [\ref{e:vsig}]) are weighted, by the column density,
\begin{eqnarray}
    \bar Z(\rb) &=& \frac{\sum_{n=1}^N N(\mathrm{H\small{I}})_n(\rb) \, (\ab_n^{\tau}(\rb)/\ab_n^{em}(\rb))}{\sum_{n=1}^N N(\mathrm{H\small{I}})_n(\rb)},\\
    v_c &=& \frac{\sum_v v \, \tilde T_b(\rb)}{\sum_v \tilde T_b(\rb)}\\ 
    \bar \sigma^2(\rb)&=&\frac{\sum_v (v-v_c(\rb))^2 \, \tilde T_b(\rb)}{\sum_v \tilde T_b(\rb)}.
\end{eqnarray}


\section*{Acknowledgements}
We thank the anonymous referee for a focused and thoughtful report. 
This work was partially supported by HST grant HST-GO-13840.008-A, and by UNC Chapel Hill. It has made use of NASA's Astrophysics Data System. Special thanks to the PSI2/Interstellar Institute Programs at Universit\'{e} Paris-Saclay in 2017--2020. The authors thank B.~Wakker and F.~J.~Lockman for detailed comments on (earlier versions of) the manuscript. Simulations were run on the UNC clusters {\tt longleaf} and {\tt dogwood}, administered by UNC's Information and Technology Services.

\section*{Data Availability Statement}
Raw and reduced simulation data are available on request. 

\bibliographystyle{mnras}
\bibliography{./references}


\bsp
\label{lastpage}
\end{document}